\documentclass[12pt,preprint,iop,appendixfloats,numberedappendix]{emulateapj}
\usepackage{amsfonts,amsmath,apjfonts}
\usepackage{natbib}
\usepackage{graphicx}
\usepackage[caption=false]{subfig}
\usepackage{booktabs}
\usepackage[breaklinks,colorlinks,citecolor=blue]{hyperref}
\usepackage[all]{hypcap}

\def\cfa{1}

\shorttitle{The Type I Superluminous Supernova PS16aqv}
\shortauthors{Blanchard et al.}

\begin{document}

\title{The Type I Superluminous Supernova PS16aqv:  Lightcurve Complexity and Deep Limits on Radioactive Ejecta in a Fast Event}

\author{
P. K. Blanchard\altaffilmark{\cfa,2}, 
M. Nicholl\altaffilmark{\cfa},
E. Berger\altaffilmark{\cfa},
R. Chornock\altaffilmark{3},
R. Margutti\altaffilmark{4},
D. Milisavljevic\altaffilmark{5},
W. Fong\altaffilmark{4},
C. MacLeod\altaffilmark{\cfa},
and K. Bhirombhakdi\altaffilmark{3}
}
\email{pblanchard@cfa.harvard.edu}

\altaffiltext{1}{Harvard-Smithsonian Center for Astrophysics, 60 Garden Street, Cambridge, MA 02138, USA}
\altaffiltext{2}{NSF GRFP Fellow}
\altaffiltext{3}{Astrophysical Institute, Department of Physics and Astronomy, 251B Clippinger Lab, Ohio University, Athens, OH 45701, USA}
\altaffiltext{4}{Center for Interdisciplinary Exploration and Research in Astrophysics (CIERA) and Department of Physics and Astronomy, Northwestern University, Evanston, IL 60208, USA}
\altaffiltext{5}{Department of Physics and Astronomy, Purdue University, 525 Northwestern Avenue, West Lafayette, IN, 47907, USA}

\begin{abstract}
\medskip

We present UV/optical photometric and spectroscopic observations of PS16aqv (SN\,2016ard), a Type I superluminous supernova (SLSN-I) classified as part of our search for low-$z$ SLSNe.  PS16aqv is most similar in timescale and spectroscopic evolution to fast evolving SLSNe-I and reached a peak absolute magnitude of $M_{r} \approx -22.1$.  The lightcurves exhibit a significant undulation at 30 rest-frame days after peak, with a behavior similar to undulations seen in the slowly fading SLSN-I SN\,2015bn.  This similarity strengthens the case that fast and slow SLSNe-I form a continuum with a common origin.  At $\approx\!80$ days after peak, the lightcurves exhibit a transition to a slow decline, followed by significant subsequent steepening, indicative of a plateau phase or a second significant undulation.  Deep limits at $\approx280$ days after peak imply a tight constraint on the nickel mass, $M_{\rm Ni} \lesssim 0.35$\! M$_{\odot}$ (lower than for previous SLSNe-I), and indicate that some SLSNe-I do not produce significantly more nickel than normal Type Ic SNe.  Using {\tt MOSFiT}, we model the lightcurve with a magnetar central engine model and find $P_{\rm spin} \approx 0.9$ ms, $B \approx 1.5 \times 10^{14}$ G, and $M_{\rm ej} \approx 16$ M$_{\odot}$.  The implied rapid spin-down time and large reservoir of available energy coupled with the high ejecta mass may account for the fast evolving lightcurve and slow spectroscopic evolution.  We also study the location of PS16aqv within its host galaxy and find that it occurred at an offset of $2.46 \pm 0.21$ kpc from the central, most active star-forming region.  We find the host galaxy exhibits low metallicity and spatially varying extinction and star formation rate, with the explosion site of PS16aqv exhibiting lower values than the central region.  The complexity seen in the lightcurves of PS16aqv and other events highlights the importance of obtaining well-sampled lightcurves for exploring deviations from a uniform decline.         

\smallskip
\end{abstract}


\section{Introduction}

Modern optical time-domain surveys, unbiased with respect to host galaxy environment, have discovered superluminous supernovae (SLSNe) with luminosities exceeding those of normal supernovae (SNe) by at least an order of magnitude \citep{Quimby2011,Chomiuk2011,Gal-Yam2012}.  This has dramatically increased the known diversity of SNe, and fueled theoretical and observational efforts to understand the most extreme ways that massive stars end their lives.  Similar to their normal luminosity counterparts, SLSNe can be divided into two classes based on the presence or absence of hydrogen emission lines in their spectra.  The majority of hydrogen-rich Type II SLSNe show narrow and intermediate width Balmer emission lines and are thus the most luminous examples of Type IIn SNe \citep[but see][]{Inserra2018}.  Their luminosities can be explained by interaction with a slow-moving circumstellar medium \citep[CSM;][]{Smith2007,ChevalierIrwin2011}.  Hydrogen-poor Type I SLSNe (hereafter SLSNe-I) are characterized at early times by blue spectra indicating temperatures of $\gtrsim\!10^{4}$ K with few features other than distinctive \ion{O}{2} absorption features at wavelengths of $\sim 3600-4600$ \AA\ \citep{Gal-Yam2012}.  As the temperature decreases, their spectra begin to resemble normal luminosity Type Ic SNe suggesting that their ejecta have similar compositions, but with an additional, persistent heating source in SLSNe-I.  

The proposed models for the power sources of SLSNe-I are a central engine \citep{KasenBildsten2010}, hydrogen-free CSM interaction \citep{ChevalierIrwin2011}, or an over-abundant production of radioactive $^{56}$Ni \citep{HegerWoosley2002,Gal-Yam2009}.  While CSM interaction can explain the lightcurves of SLSNe-I \citep{Chatzopoulos2013,Nicholl2014} and the emergence of late-time H$\alpha$ emission in some events suggests eventual interaction with material at $\sim\!10^{16}$ cm from the progenitor \citep{Yan2017}, there is no spectroscopic evidence that CSM interaction is the dominant power source near peak.  Central engine models, the most popular being the spin-down of a rapidly rotating magnetar \citep{KasenBildsten2010}, are also able to reproduce the lightcurves of SLSNe-I \citep{Inserra2013,Nicholl2014,Nicholl2017}.  In addition, the early phase spectra of SLSNe-I have generally favored spectroscopic models produced by a central, illuminating source rather than pair-instability models in which a significant amount of $^{56}$Ni is produced \citep{Dessart2012,Mazzali2016}.  This appears to also hold true with the few SLSNe-I that have nebular phase spectra, which have indicated similar ejecta compositions and velocity structures with SNe associated with long gamma-ray bursts \citep[GRBs;][]{Milisavljevic2013,Nicholl2016b,Jerkstrand2016,Jerkstrand2017}.  Furthermore, host galaxy studies of SLSNe-I have shown that they occur in metal-poor dwarf galaxies \citep{Chen2013,Lunnan2014,Leloudas2015,Perley2016}, similar to long GRB hosts, and radio and X-ray observations of SLSNe-I indicate low-density circumstellar environments \citep{Nicholl2016,Margutti2017,Coppejans2018}, lower than expected if CSM interaction is the dominant power source.  

Given the lines of evidence favoring the magnetar central engine model for SLSNe-I, it is important to study whether this model can explain the full range of SLSN-I properties, given that this class exhibits a wide range of photometric behavior.  This observed diversity, notably the order of magnitude spread in rise and decline timescales \citep{Nicholl2015} and peak bolometric luminosity, has led to debate in the literature regarding whether SLSNe-I constitute a single class resulting from a single physical mechanism with varying parameters or if sub-classes exist which reflect the presence of multiple power sources and/or explosion mechanisms.  Modeling of large samples of SLSN-I lightcurves has suggested that a continuum of ejecta and engine properties can account for the range of known SLSNe-I \citep{Nicholl2017}. However, \citet{Inserra2018a} found that slower SLSNe-I have a shallower velocity gradient, keeping open the possibility that some significant physical differences may exist among SLSNe-I. 

One important diagnostic is the presence of short timescale lightcurve variability, often referred to as undulations.  This was first noted for the slowly evolving SN\,2015bn \citep{Nicholl2016}. \citet{Inserra2017} found similar undulations in other slow SLSNe-I, but such behavior was difficult to detect in faster evolving events, at least in part due to the steeper overall lightcurves and shorter sampling baseline.  \citet{Nicholl2014} identified one fast SLSN-I, SSS120810, that did show significant variability.  High-amplitude lightcurve undulations have also been found in iPTF13dcc \citep{Vreeswijk2017} and iPTF15esb \citep{Yan2017}. 

Recently, we began a targeted search for low-$z$ SLSNe which can be studied in detail near peak and to late times.  Here we present observations of PS16aqv, a SLSN-I at $z = 0.2025$ discovered as part of this search.  Through our extensive follow-up campaign we were able to obtain well-sampled lightcurves and spectra and we find that PS16aqv is overall most similar to the fast evolving SLSNe-I.  However, PS16aqv stands out as a fast declining event with clear evidence for undulations in its lightcurve, remarkably similar to SN\,2015bn \citep{Nicholl2016}.  Furthermore, the lightcurve exhibited a transition to a very slow decline phase followed by rapid fading, indicating complex behavior at late-times.  

The paper is structured as follows.  In Section 2 we present our photometric and spectroscopic data of PS16aqv.  In Section 3 we present the observational characteristics of PS16aqv with comparisons to other SLSNe-I.  In Section 4 we discuss our MCMC modeling of the lightcurve with a magnetar model.  In Section 5 we analyze the properties of the host galaxy and environment in which PS16aqv occured.  In Section 6 we discuss the implications of and possible physical mechanisms responsible for the lightcurve undulations in PS16aqv, their similarity to those in other events, the late-time deviations from a smooth decline, and limits on the nickel mass.  Finally, we conclude in Section 7. 

In this paper we use $H_{0} = 67$ km s$^{-1}$ Mpc$^{-1}$, $\Omega_{m} = 0.32$, and $\Omega_{\Lambda} = 0.68$ \citep{Planck2013}, resulting in a luminosity distance of 1035 Mpc to PS16aqv.  The Galactic extinction along the line of sight to PS16aqv is $E(B-V) = 0.0433 \pm 0.0011$ mag \citep{SF2011}.

\begin{figure*}[t!]
\begin{center}
\includegraphics[scale=0.47]{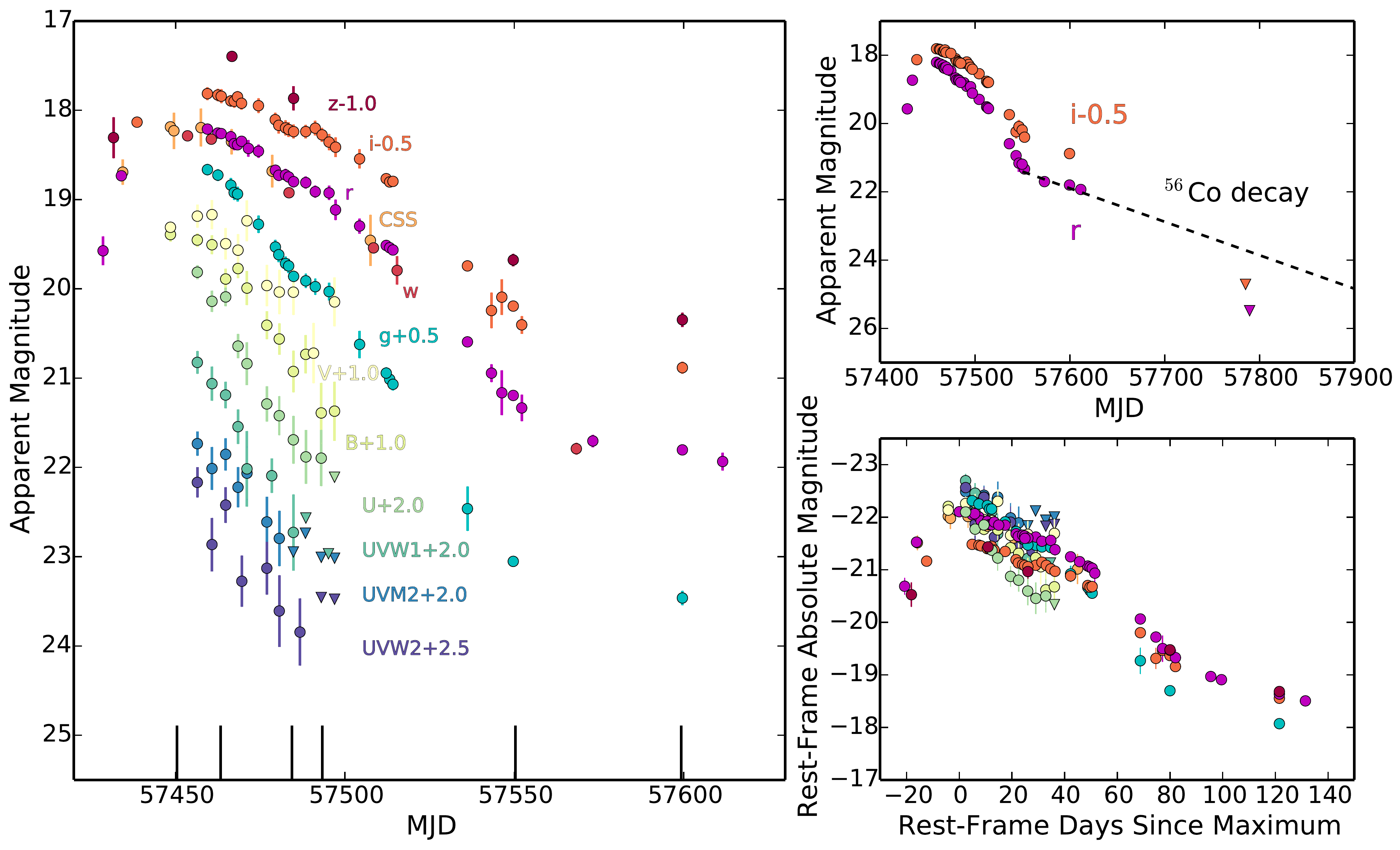}
\end{center}
\caption{Left: UV and optical lightcurves of PS16aqv corrected for Galactic extinction with band offsets for clarity.  The \textit{Swift}/UVOT filters $U$, $B$, $V$, $UVW1$, $UVM2$, and $UVW2$ and CSS data are in Vega magnitudes and all others are in AB magnitudes.  Vertical lines at the bottom indicate the epochs of our spectra.  The good time-sampling clearly reveals interesting behavior such as lightcurve undulations.  Top Right: The $r$-band and $i$-band lightcurves of PS16aqv including deep upper limits at $\approx 280$ rest-frame days after peak.  While the decline rate closely matches that for fully-trapped $^{56}$Co decay at $\sim 80 - 130$ rest-frame days after peak, the decline rate must dramatically increase at later times to account for the upper limits.  Bottom Right: Rest-frame absolute magnitude lightcurves (no offsets), which take into account K-corrections (measured from our spectra), Galactic extinction, and internal host galaxy extinction inferred from our lightcurve modeling (Section \ref{sec:mag}).  PS16aqv exhibited a peak absolute $r$-band magnitude of $M_{r} = -22.10\pm0.12$.}
\label{obsLC}
\end{figure*}

\capstartfalse
\begin{deluxetable*}{ccccccc}[!htb]
\tablecolumns{7}
\tabcolsep0.1in\footnotesize
\tablewidth{7in}
\tablecaption{Spectroscopic Observations of PS16aqv  
\label{tab:spec}}
\tablehead {
\colhead {Date}   &
\colhead {MJD}     &
\colhead {Phase\tablenotemark{a}} &
\colhead {Telescope} &
\colhead{Instrument}  &
\colhead {Airmass}   &
\colhead {Resolution (\AA)}           
}   
\startdata
2 March 2016 & 57450.4 & $-$2.5  & MDM/Hiltner & OSMOS & 1.53 & 5 \\
15 March 2016 & 57463.3 & +8.2  & FLWO 60-inch & FAST & 1.49 & 5.7 \\
5 April 2016 & 57484.3 & +25.7  & Magellan/Baade & IMACS & 1.17 & 5.4 \\
14 April 2016 & 57493.3 & +33.2  & MMT & Blue Channel & 1.45 & 4 \\
10 June 2016 & 57550.3 & +80.6 & Magellan/Clay & LDSS3c & 1.06 & 7.5 \\
29 July 2016 & 57599.3 & +121.3 & Magellan/Clay & LDSS3c & 1.28 & 7.5 \\
26 January 2017 & 57780.3 & +271.9 & Gemini-N & GMOS & 1.26  & 11   
\enddata
\tablenotetext{a}{Rest-frame days since peak bolometric brightness}
\end{deluxetable*}   
\capstarttrue

\begin{figure*}[t!]
\begin{center}
\includegraphics[scale=0.46]{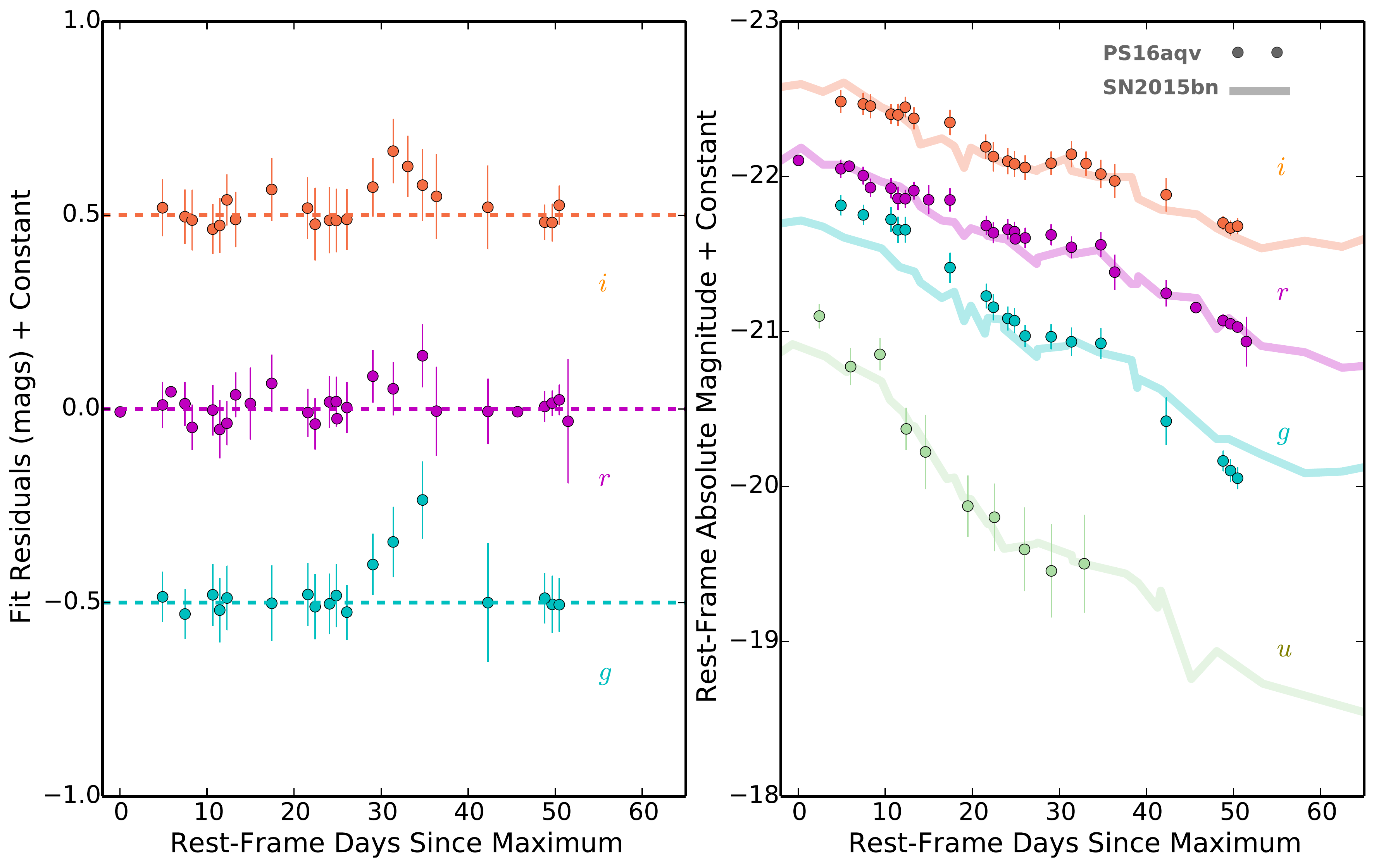}
\end{center}
\caption{Left: Residuals from low-order polynomial fits to the post-peak rest-frame $gri$ lightcurves of PS16aqv.  At $\approx30$ days after peak the lightcurves show a ``knee'', or undulation, lasting for about 10 days which appears to have a slightly higher amplitude in $g$-band.  Such a feature has been seen in SN\,2015bn at $\approx50$ days after peak \citep{Nicholl2016}.  Right: Rest-frame $ugri$ lightcurves of PS16aqv compared to SN\,2015bn after compressing the SN\,2015bn lightcurves in time by 40\%.  This time compression highlights the similar amplitudes of the lightcurve undulations in the two events.  While the undulations occur on different timescales, as might be expected due to the overall lightcurve timescale difference, the behavior is similar.}
\label{LCpoly}
\end{figure*}

\section{Observations of PS16aqv}

\subsection{Discovery}

PS16aqv, also known as SN\,2016ard, was classified as part of our program to identify SLSNe from the Pan-STARRS Search for Transients \citep[PSST;][]{Huber2015}, which publicly reports stationary transients from the on-going Pan-STARRS near-Earth object survey.  PS16aqv was first detected by PSST on 10 February 2016 but due to being partially located in a detector chip gap it was not flagged by the PSST detection software until 20 February 2016 when it reached a magnitude of $i \approx 18.7$.  The SN was independently discovered by the Catalina Real-Time Transient Survey \citep[CRTS;][]{Drake2009} on 16 February 2016 and was designated CSS160216:141045-100935.  Examining the associated Pan-STARRS 3$\pi$ deep stack, we found a marginal detection of a host galaxy with $r \approx 22.6$ mag.  The large brightness contrast between the transient and host galaxy motivated us to initiate follow-up observations.  A spectrum obtained on 2 March 2016 using the Ohio State Multiple Object Spectrograph \citep[OSMOS;][]{Martini2011} on the 2.4-m Hiltner telescope at MDM Observatory exhibited a blue continuum with weak spectral features consistent with the \ion{O}{2} lines commonly seen in the pre- and near-maximum light spectra of SLSNe-I.  The redshift of $z\approx0.20$ implied by this identification (later confirmed from host galaxy emission lines to be $z = 0.2025 \pm 0.0003$) yielded an absolute magnitude for the PSST detection of $M_{i}\approx -21.3$, confirming the superluminous nature of the event.  Following classification we obtained additional photometric and spectroscopic observations.       

\subsection{UV and Optical Photometry}
\label{sec:optobs}

We obtained images of PS16aqv in the $BVR$ filters on 1 March 2016 using the 1.3-m telescope at MDM Observatory and in the $gri$ filters using the 48-inch telescope at the Fred Lawrence Whipple Observatory (FLWO) from 11 March to 3 July 2016.  We also obtained images using IMACS \citep{Dressler2011} and LDSS3c \citep{Stevenson2016} on the Magellan 6.5-m telescopes at Las Campanas Observatory in the $griz$ filters extending to 19 March 2018.  We reduced the images using standard techniques and performed photometry using point-spread function (PSF) fitting implemented with the {\tt daophot} IRAF package.  Instrumental magnitudes in the $griz$ filters were calibrated to the Pan-STARRS 3$\pi$ photometric system in AB magnitudes using zeropoints calculated from field comparison stars.  The $BVR$ instrumental magnitudes were calibrated to Vega magnitudes using Landolt fields observed on the same night.  The uncertainties on the calibrated magnitudes include the uncertainty resulting from the PSF fit and the uncertainty on the nightly zeropoints.  We also obtained observations of PS16aqv using the UV/Optical Telescope \citep[UVOT;][]{Roming05} onboard the \textit{Swift} satellite in the $U$, $B$, $V$, $UVW1$, $UVM2$, and $UVW2$ filters.  We analyzed the data following the prescription of \citet{Brown09} using the updated calibration files and zeropoints from \citet{Breeveld11}.  PS16aqv was detected in 11 epochs from 9 March to 18 April 2016.    

From discovery to about two months after peak brightness, the flux in our images is dominated by that of the SN in all filters and therefore host subtraction is not necessary.  However, as the SN faded, the host contribution became significant and required careful host subtraction to isolate the SN flux.  We performed image subtraction using {\tt HOTPANTS} \citep{Becker2015} on our $griz$ images obtained after the gap in observations around 15 May 2016 (MJD 57523).  For $g$, $r$, and $z$ observations after this date, we use deep templates obtained on 17 July 2017 with IMACS and for $i$-band observations we use a deep template obtained on 19 March 2018 with LDSS3c.  Subtracting these templates from similarly deep $i$- and $r$-band images taken on 31 January and 2 February 2017 ($\sim\!$10 months after peak brightness), respectively, we find no detectable SN flux indicating PS16aqv had already faded significantly by early 2017.  We measure upper limits on the brightness of PS16aqv in the 31 January and 2 February 2017 images using the following procedure.  We inject a fake point source at the position of PS16aqv (measured using relative astronomy with images containing SN flux) and then we perform image subtraction using the templates.  We repeat this for a range of magnitudes and consider the 3$\sigma$ upper limit to correspond to a source detected at 3$\sigma$ in the subtracted image.  We find an upper limit of $r>25.6$ mag from the 2 February 2017 image and $i>25.3$ mag from the 31 January 2017 image.     

PS16aqv was also detected in several epochs by PSST in the $w$, $r$, $i$, and $z$ filters, as well as by the unfiltered CRTS.  The earliest two PSST detections were not recorded by the PSST pipeline due to proximity to a chip gap, but by examining the 2D frames and performing PSF fitting photometry we were able to recover the flux.  For the purpose of calculating the rest-frame lightcurves we converted the $w$-band magnitudes to $r$-band using a shift of $-0.13$ mag empirically determined from the lightcurves.  No correction was applied to the CRTS data as they are already well-matched to our $r$-band measurements.        

Our ground-based photometry, in addition to the PSST and CRTS data, is listed in Table \ref{tab:ground} and the \textit{Swift}/UVOT photometry is listed in Table \ref{tab:swift}.  In Figure \ref{obsLC} we show the corresponding lightcurves.

\subsubsection{\textit{Hubble Space Telescope (HST)} Observations}
We obtained \textit{HST} observations of PS16aqv on 27 December 2017 using the Advanced Camera for Surveys (ACS) Wide Field Camera (WFC) with the F775W filter (PID: 15162; PI: Blanchard).  Four dithered images were corrected for optical distortion and drizzle-combined to a finer grid (0.035'' per pixel) using the {\tt astrodrizzle} task in the {\tt drizzlepac}\footnote{\url{http://drizzlepac.stsci.edu/}} software package provided by STScI.  We examine the location of PS16aqv, which we determined by performing relative astrometry with an LDSS3c image containing the transient, and find no point source is detected at the measured position.  However, the precision on the position is sufficient to yield information on the environment of PS16aqv (Section \ref{sec:host}).

\subsection{Optical Spectroscopy}
We obtained 7 epochs of spectroscopy of PS16aqv spanning $-2.5$ to $+272$ rest-frame days since maximum brightness using OSMOS on the 2.4-m Hiltner telescope, the FAST Spectrograph \citep{Fabricant1998} on the 60-inch telescope at FLWO, IMACS and LDSS3c on the Magellan 6.5-m telescopes, the Blue Channel Spectrograph \citep{Schmidt1989} on the 6.5-m MMT telescope, and GMOS-N on the 8-m Gemini-North telescope.  The observation epochs, airmasses, and spectral resolutions are given in Table \ref{tab:spec}.  The 2D spectra were reduced using standard techniques in IRAF to extract 1D wavelength-calibrated spectra.  Relative flux calibration was achieved using standard stars observed on the same nights.  If needed, the spectra were scaled to match contemporaneous photometry to achieve an absolute flux calibration.  The spectra were corrected for Galactic extinction and transformed to the rest-frame of PS16aqv for analysis.     

\subsection{X-ray Observations}
We obtained X-ray observations of PS16aqv using the X-ray Telescope \citep[XRT;][]{Burrows2005} onboard \textit{Swift} from 9 March to 10 June 2016.  The data analysis and results are provided in \citet{Margutti2017}.  We find no detection of an X-ray source at the position of PS16aqv in any epoch, resulting in a combined unabsorbed flux upper limit of $F_{X} < 1.5 \times 10^{-14}$ erg s$^{-1}$ cm$^{-2}$ ($0.3 - 10$ keV).  

\begin{figure*}[t!]
\begin{center}
\includegraphics[scale=0.397]{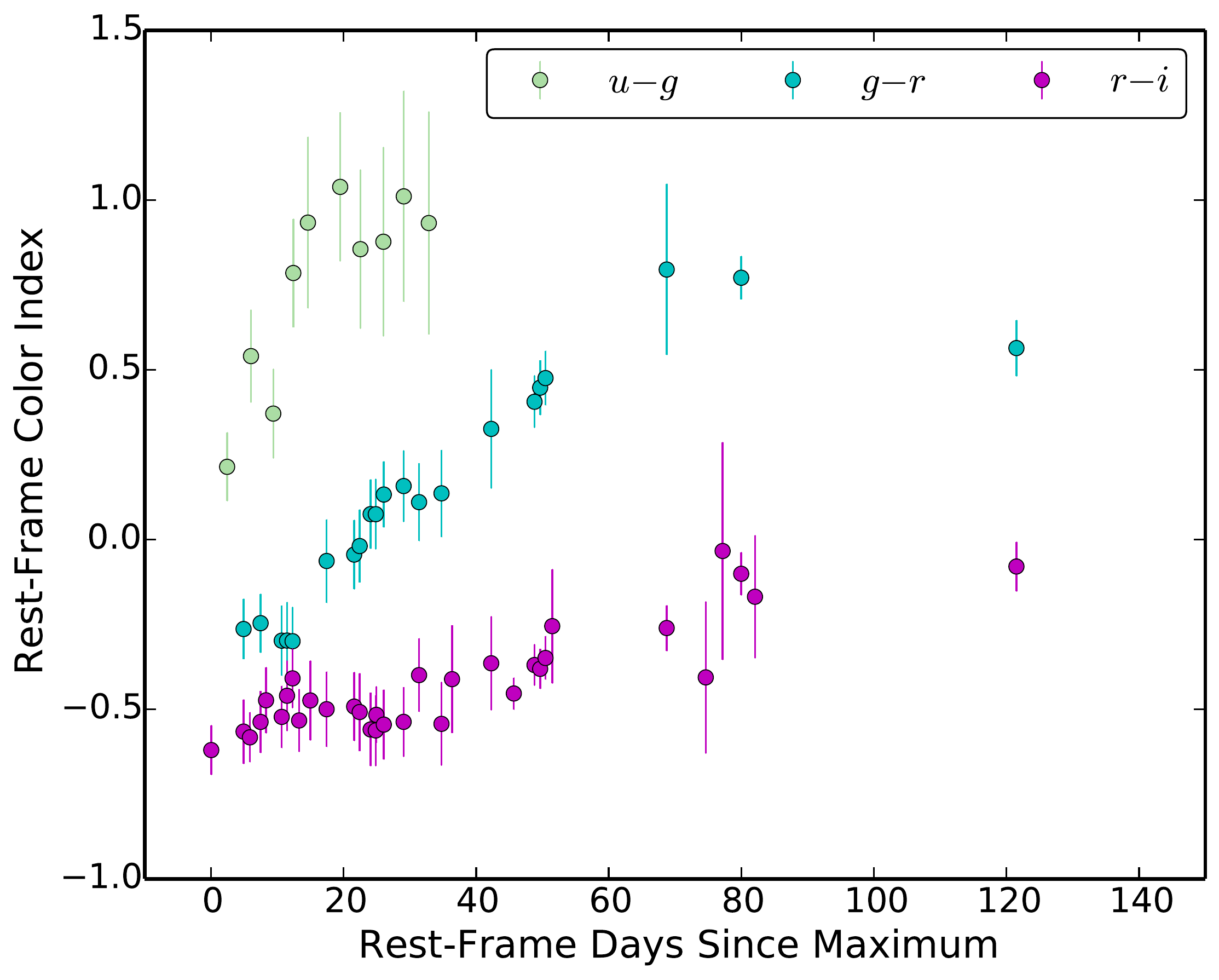}
\includegraphics[scale=0.397]{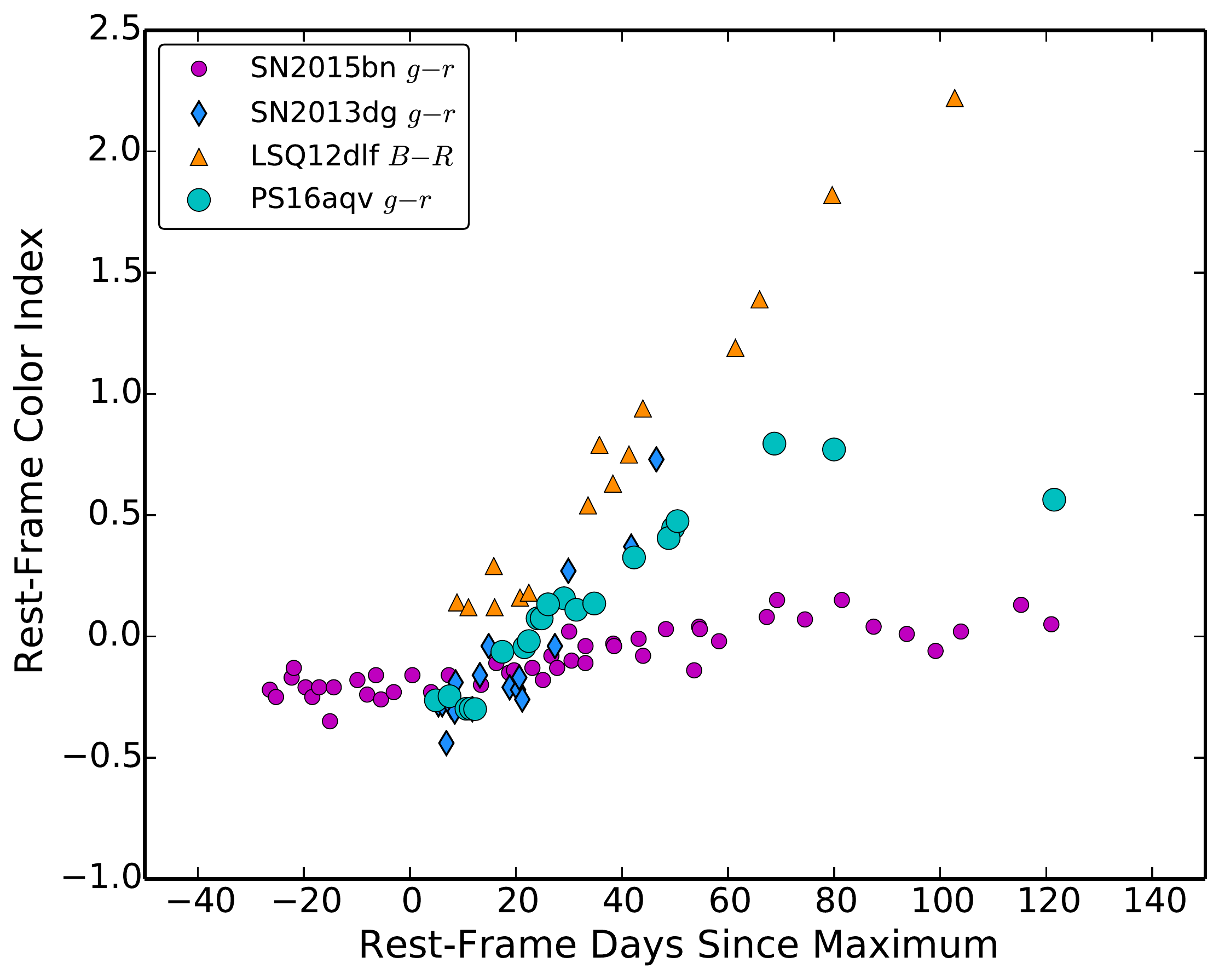}
\end{center}
\caption{Left: Rest-frame color evolution of PS16aqv in the $u-g$, $g-r$, and $r-i$ colors.  Right: Rest-frame $g-r$ (or $B-R$) color evolution of PS16aqv compared to SN\,2015bn, SN\,2013dg, and LSQ12dlf \citep{Nicholl2014,Nicholl2016}.  While all of these events have blue $g-r$ colors near peak brightness, they evolve at different rates.  Commensurate with the overall lightcurve timescale differences, PS16aqv exhibits a faster color evolution than SN\,2015bn and is slower than SN\,2013dg and LSQ12dlf.  The colors of PS16aqv and SN\,2015bn appear to reach a plateau value at about $+80$ days whereas LSQ12dlf continues a consistent reddening with time.}
\label{colors}
\end{figure*}

\begin{figure}[t!]
\begin{center}
\includegraphics[scale=0.35]{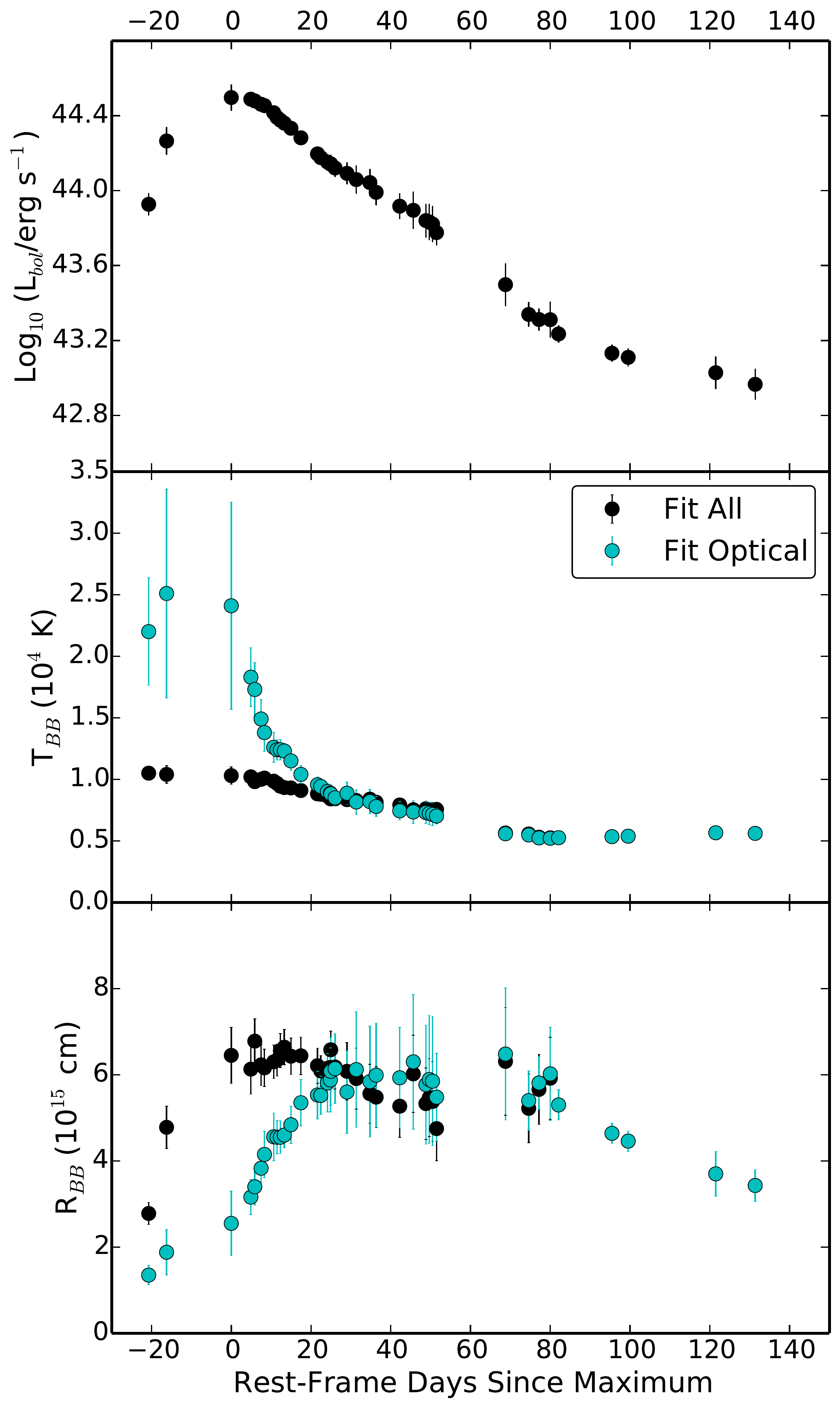}
\end{center}
\caption{Top: Bolometric lightcurve of PS16aqv.  Middle:  Temperature evolution inferred from the blackbody fits to each epoch.  We show both fits to the entire SED and the optical data only.  The earliest temperature points rely on extrapolation due to the lack of good data on the rise and are therefore very uncertain.  Bottom: Photospheric radius inferred from the blackbody fits.  A single blackbody yields a poor fit to the entire SED due to UV absorption and so we consider the temperatures and radii inferred from the optical-only fits to be a better representation of the photosphere.  As seen in other SLSNe-I, the temperature reaches a constant value and the photosphere begins receding into the ejecta.}
\label{bolLC}
\end{figure}

\section{Observational Characteristics of PS16aqv}

\subsection{Multi-Band Observed and Rest-Frame Light Curves}
\label{sec:LC}

We present the observed UV/optical lightcurves of PS16aqv in Figure \ref{obsLC}.  Following the earliest observation by PSST, PS16aqv brightened by about 1.4 magnitudes in 25 days to maximum brightness, a longer rise than most normal Type Ic SNe and consistent with SLSNe-I \citep{Nicholl2015}.  Upon reaching maximum brightness PS16aqv mirrored its slow rise with a slow decline in $r$- and $i$-band, and with a faster decline rate in $g$ and bluer filters.  About 30 days after maximum brightness, the decline rate of PS16aqv slows considerably in $g$-, $r$-, and $i$-band, forming a prominent ``knee'' \citep[using the terminology of][]{Nicholl2016} in the lightcurves. There is also evidence of this knee in $u$ and perhaps bluer bands, though the large error bars on the latest UV points makes this unclear.  Following the knee, the decline rate approximately resumes the same rate in $g$-band and a slightly higher rate in $r$- and $i$-band, until about 100 observer-frame days after peak where the $griz$ lightcurves begin to show a clear flattening.  The observed decline rate in $r$-band at this phase is about 0.008 mag/day, roughly matching the decline rate due to fully-trapped $^{56}$Co decay powering.  Extrapolating this slow decline to the epoch of our $r$- and $i$-band upper limits at about 330 observer-frame days after peak, we find this slow phase is not sustained and that PS16aqv must have resumed a faster decline to account for the upper limits.     

We calculate the rest-frame absolute magnitudes in each filter using the precise redshift of PS16aqv with a correction for Galactic extinction and K-corrections.  We assume an internal host extinction of $A_{V} = 0.55$ mag based on our lightcurve modeling in Section \ref{sec:mag}.  The K-corrections were determined from our observed spectra by convolving each filter bandpass with the observer- and rest-frame spectra using the K-correction code {\tt SNAKE} \citep{Inserra2018}.  We then fit a polynomial to the set of K-corrections as a function of time in each filter, allowing us to estimate the K-correction at each photometric epoch.  Due to the lack of NUV spectroscopic coverage, the NUV K-corrections rely on blackbody fits to the optical spectra and are thus only approximate.  In Figure \ref{obsLC} we show the resulting rest-frame lightcurves of PS16aqv spanning a total range of about $-20$ to $+130$ days relative to peak brightness, showing that PS16aqv reached a maximum luminosity of $M_{r} = -22.10\pm0.12$.            

While occurring on a different timescale, the knee observed in PS16aqv at 30 days after peak is similar to that seen in SN\,2015bn at 50 days after peak \citep{Nicholl2016}.  To help visualize the knee in PS16aqv, also termed an undulation, we fit low-order polynomials to the post-peak $gri$ lightcurves to remove the overall decline trend.  In Figure \ref{LCpoly} we show the residuals of these fits, which show the undulations are coherent in time across multiple filters and lasted for about 10 days.  The amplitudes of the undulations in each filter are the same as those in SN\,2015bn and there is a slightly higher amplitude in $g$-band.  To test the significance of the undulations in PS16aqv, we perform a runs test on the residuals in each filter.  We find that the number of runs in each filter shows a statistically significant deviation from the expected number, indicating the residuals are not completely random.  In Figure \ref{LCpoly} we also show the $ugri$ rest-frame lightcurves of PS16aqv compared to the rest-frame lightcurves of SN\,2015bn compressed in time by 40\% to match the observed lightcurve knees in the two events.  The resulting lightcurves show a striking similarity from maximum brightness to about 50 days after.        

In Figure \ref{colors} we show PS16aqv's rest-frame color evolution in the $u-g$, $g-r$, and $r-i$ color indices as well as a comparison of the $g-r$ color evolution with that of several other SLSNe-I.  The color evolution of PS16aqv is slowest in the $r-i$ color, taking several months to redden by half a magnitude, with progressively faster reddening in $g-r$ and $u-g$.  This is because $g-r$ and $u-g$ probe the peak of the thermal continuum, whereas $r-i$ is on the Rayleigh-Jeans tail.  After evolving steadily for about 80 days, the $r-i$ and $g-r$ colors appear to show little evolution between 80 and 120 days after maximum brightness.  Extrapolating the $g-r$ evolution to peak we find $g-r \approx -0.45$, bluer than SN\,2015bn at peak but similar to the extrapolation of SN\,2013dg.  From maximum brightness to about 80 days later, the $g-r$ color of PS16aqv clearly reddens at a faster rate than that of SN\,2015bn, as expected from the overall faster lightcurve evolution of PS16aqv.  However, they both show a flattening in the $g-r$ color evolution after about 80 days, which is not seen in LSQ12dlf.  In addition, the $g-r$ color evolution of PS16aqv is slower than that of LSQ12dlf and SN\,2013dg and therefore shows an intermediate color evolution.

\begin{figure}[t!]
\begin{center}
\includegraphics[scale=0.37]{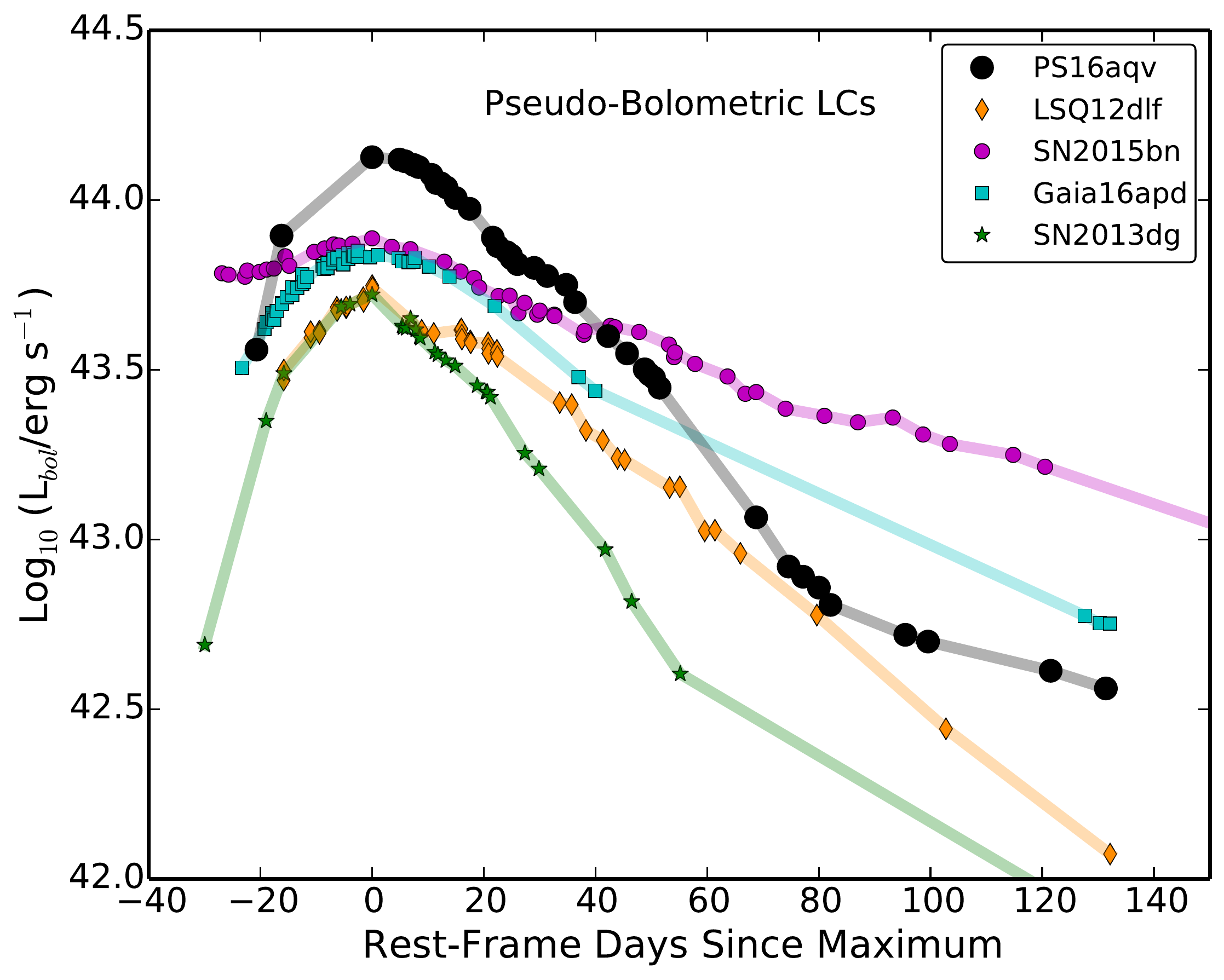}
\end{center}
\caption{Pseudo-bolometric lightcurves calculated from $griz$ (or $BVRI$) observations for PS16aqv, LSQ12dlf, SN\,2015bn, Gaia16apd, and SN\,2013dg \citep{Nicholl2014,Nicholl2016,Nicholl2017a}.  PS16aqv shows a significant flattening in its decline rate around $+80$ days, which is not seen in LSQ12dlf or SN\,2013dg.}
\label{bolLCcomp}
\end{figure}

\begin{figure*}[t!]
\begin{center}
\includegraphics[scale=0.5]{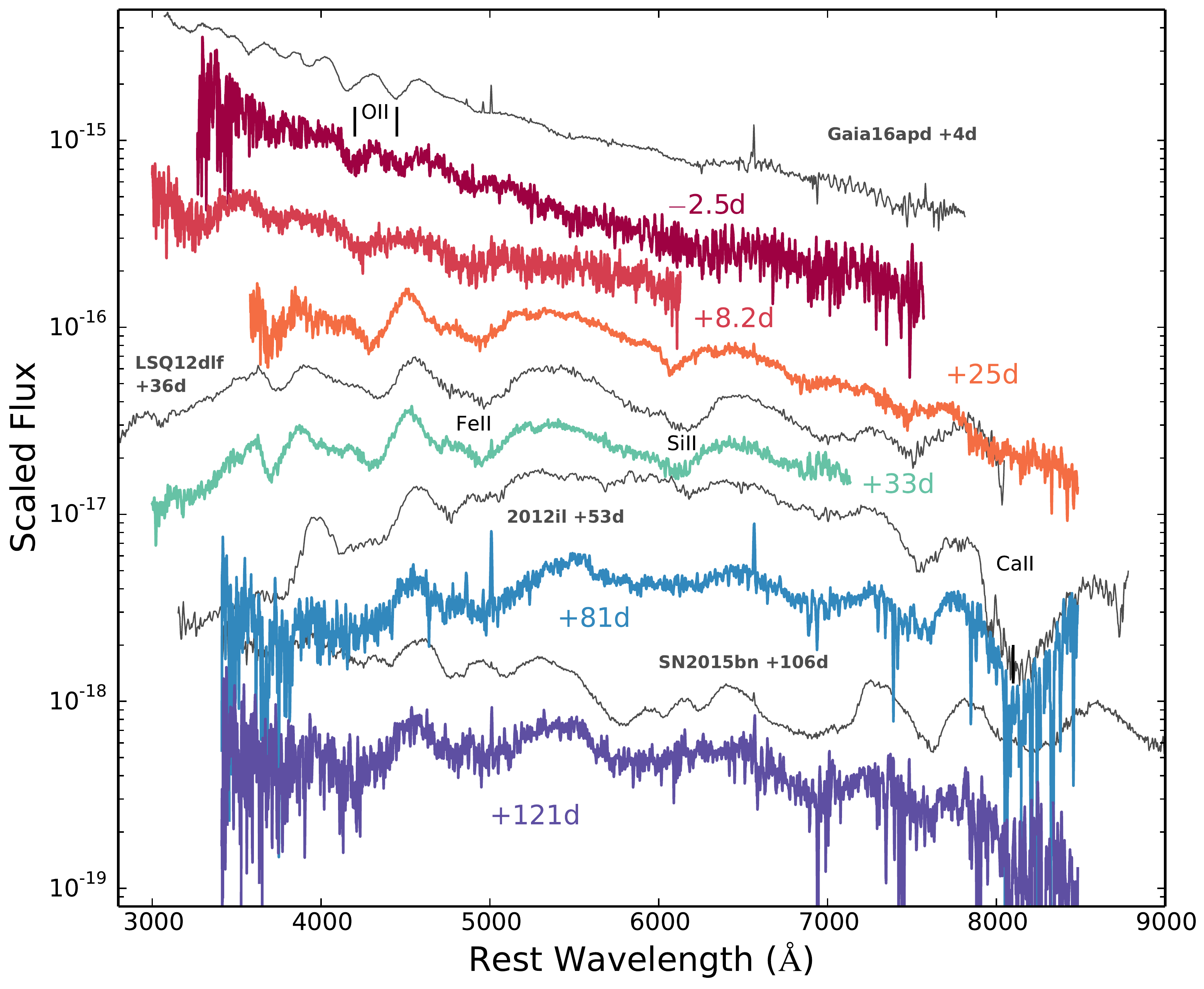}
\end{center}
\caption{Spectra of PS16aqv from $-2.5$ to $+121$ rest-frame days after peak bolometric brightness (colored spectra; phases marked) and comparisons with Gaia16apd, LSQ12dlf, SN\,2012il, and SN\,2015bn \citep[grey spectra;][]{Inserra2013,Nicholl2014,Nicholl2016,Nicholl2017a}.  PS16aqv exhibits the typical early blue continuum and \ion{O}{2} absorption lines seen in SLSNe-I and subsequent development of lower ionization lines as the ejecta cool.  At $+81$ and $+121$ days, PS16aqv does not show definitive nebular emission lines like SN\,2015bn at $+106$ days after peak, indicating a slow spectroscopic evolution.  Host emission lines are detected in the $+81$ day spectrum from which we measured a redshift of $z = 0.2025 \pm 0.0003$.}
\label{spec}
\end{figure*}

\subsection{Bolometric Lightcurve}

To understand the total energy output of PS16aqv, we calculate its bolometric lightcurve.  This is accomplished by integrating the rest-frame spectral energy distribution (SED) at each epoch with an $r$-band measurement, since $r$-band is the best-sampled filter.  To calculate the SED at each epoch, we interpolate the lightcurves of the other filters and if necessary, extrapolate assuming constant colors.  Most $gri$ measurements were taken on the same night.  While extrapolation is the only way to estimate the UV portion of the SED beyond $\sim40$ days, by this phase most of the flux is captured by $gri$ and so the method of extrapolation has a negligible effect on the bolometric luminosity.  To estimate the flux contribution coming from wavelengths blueward and redward of $uvw2$ and $z$, respectively, we fit separate blackbodies to the UV and optical measurements.  Due to metal line blanketing in the UV, a single blackbody does not accurately capture the full UV/optical SED.  As the SED peaks near $U$-band, the flux contribution from regions outside the observed wavelength range is a small correction.  The final bolometric luminosity estimate at each epoch therefore comes from a sum of the measured rest-frame fluxes and the estimated flux contribution from outside our observed wavelength range.     

We show the resulting bolometric lightcurve of PS16aqv in Figure \ref{bolLC}, showing that at maximum brightness it reached a bolometric luminosity of $\approx\!3.1 \times 10^{44}$ erg s$^{-1}$.  Integrating the bolometric lightcurve we find that PS16aqv radiated a total of $\approx\!1.3 \times 10^{51} $ erg over $\sim\!150$ days.  This is comparable to the total kinetic energy of typical core-collapse SNe.  We also show the blackbody temperature and photospheric radius inferred from blackbody fits to all bands and fits to the optical bands only.  Due to line blanketing in the UV we consider the temperature inferred from the fits to the optical data only to be the most reliable estimate of the photospheric temperature.  We find that near peak light $T_{\rm BB} \approx 20,000$ K and then begins a rapid decline, taking about $\sim$20 days to reach $\sim10,000$ K.  The rate of temperature decline subsequently slows down until leveling off at $\sim$5000 K at about $+80$ days.  The photospheric radius, as inferred from the optical fits, starts near $2 \times 10^{15}$ cm, reaches a maximum of about $6 \times 10^{15}$ cm (consistent with an expansion velocity of $\sim\!10^{4}$ km s$^{-1}$), and then slowly declines as the photosphere begins to recede. 

To facilitate a comparison of the bolometric lightcurve of PS16aqv with other SLSNe-I with varying levels of photometric coverage, we also calculate a pseudo $griz$ bolometric lightcurve resulting from a sum of only the $griz$ measurements.  In Figure \ref{bolLCcomp} we show a comparison of PS16aqv's pseudo-bolometric lightcurve with that of SN\,2015bn, Gaia16apd, LSQ12dlf, and SN\,2013dg \citep{Nicholl2014,Nicholl2016,Nicholl2017a}.  The timescale of the bolometric evolution of PS16aqv is generally similar to LSQ12dlf and SN\,2013dg, but the better time sampling of PS16aqv reveals a complex behavior with several changes in the decline rate.  There is a hint that LSQ12dlf may also show a lightcurve undulation about 15 days earlier than PS16aqv, further highlighting the importance of good time sampling.  Notably, PS16aqv shows an abrupt transition to a slow decline phase at $+80$ rest-frame days after peak.  This flattening corresponds to when the $g-r$ color evolution reaches a plateau (see Figure \ref{colors}) and when the temperature inferred from the blackbody fits to the optical data reaches a constant value (see Figure \ref{bolLC}).  While transitions to slow decline phases have been seen in some other fast evolving SLSNe-I \citep[e.g.~SN\,2011ke;][]{Inserra2013}, the deep late-time upper limits shown in Figure \ref{obsLC} indicate this flattening is not sustained in PS16aqv and that a second transition must have occurred.

\begin{figure*}
\begin{center}
\includegraphics[scale=0.4]{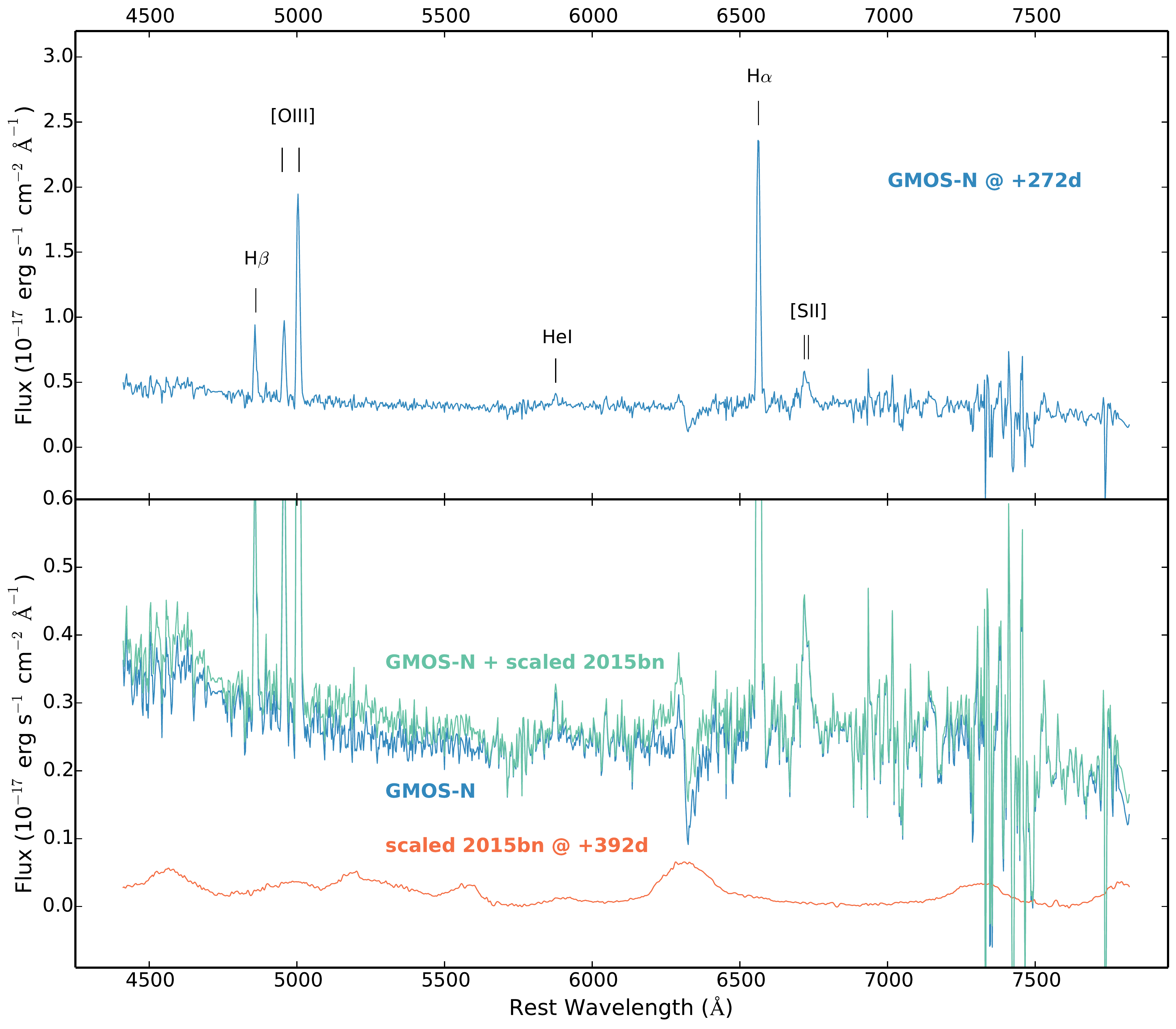}
\includegraphics[scale=0.50]{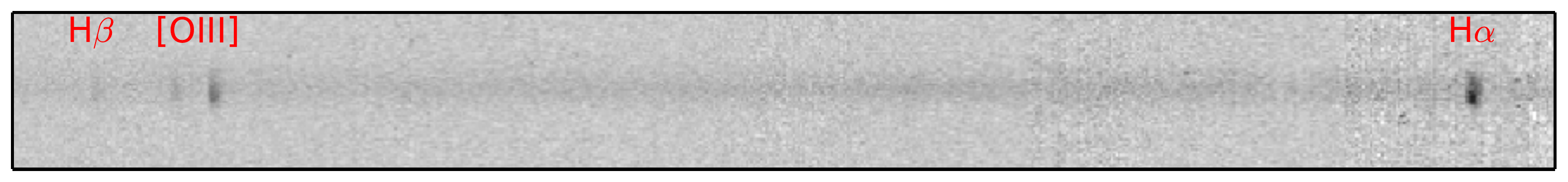}
\end{center}
\caption{Top: Gemini spectrum obtained at $+272$ rest-frame days after maximum brightness.  The spectrum lacks supernova features and is dominated by host galaxy emission lines.  Middle:  Spectrum of SN\,2015bn at $+392$ rest-frame days (lower spectrum) scaled to the upper limit on the brightness of PS16aqv at the epoch of the Gemini spectrum and the resulting spectrum after summing the Gemini and scaled SN\,2015bn spectra.  At the flux level of the upper limit, the nebular features present in SN\,2015bn are not easily discernible from the host galaxy light.  PS16aqv may have had weak nebular lines or was much fainter than the upper limit.  Bottom: 2D spectrum from which the 1D spectrum shown in the top panel was extracted.  The slit was oriented along the major-axis of the galaxy yielding spatially resolved emission line information.  The gradient of the H$\alpha$ emission line flux indicates a gradient in SFR along the galaxy.  PS16aqv occurred in a region with a relatively low SFR compared to the bright central region.}
\label{gemspec}
\end{figure*}

\subsection{Spectroscopic Evolution}
\label{sec:spec}

In Figure \ref{spec} we show the spectroscopic sequence of PS16aqv from $-2.5$ to $+121$ rest-frame days relative to peak.  For comparison, we also show spectra of Gaia16apd (\citealt{Nicholl2017a}; see also \citealt{Yan2017a}, \citealt{Kangas2017}), LSQ12dlf \citep{Nicholl2014}, SN\,2012il \citep{Inserra2013}, and SN\,2015bn \citep{Nicholl2016} at various phases.  We find that PS16aqv exhibits a similar spectroscopic evolution as previous fast evolving SLSNe-I.  The characteristic \ion{O}{2} lines are clearly detected in the $-2.5$ day spectrum.  About 10 days later the spectrum already shows signs of evolution, with a cooler continuum and weakening \ion{O}{2} lines.  By about 3 weeks after maximum brightness, the spectrum has cooled significantly and the spectral features resulting from highly ionized species such as \ion{O}{2} have given way to low ionization species such as \ion{Fe}{2}, \ion{Mg}{2}, and \ion{Si}{2}.  The spectrum of PS16aqv maintains a similar shape and shows the same spectral features for at least 10 days.  The transition from high to low ionization spectral features is typical of SLSNe-I.  

Over the next 50 days the spectrum continues to cool and shows the development of \ion{Ca}{2} absorption and possibly a hint of the emergence of [\ion{Ca}{2}] $\lambda$7300 emission.  In addition, \ion{Mg}{1}] $\lambda$4571 may also be present in PS16aqv, though we note that its coincidence with a gap in the iron opacity complicates its identification.  Moreover, the lack of other strong nebular features at this phase indicates that \ion{Mg}{1}] is unlikely the dominant source of the spectral peak near 4500 \AA, though it is clearly present in the nebular spectra of other events \citep{Nicholl2016b,Inserra2017}.  The spectrum shows little change from $+81$ to $+121$ days after maximum brightness.  In addition, these two later epochs show narrow host emission lines indicating some host contamination.  The $+81$ and $+121$ day spectra are dominated by photospheric features and do not show strong nebular lines, surprising given PS16aqv's fast lightcurve evolution.  In contrast, SN\,2015bn already shows a strong [\ion{Ca}{2}] $\lambda$7300 emission line at $+106$ days.  This has also been seen in other slowly evolving SLSNe-I (appearing as early as $+50$ days) and may be due to different emitting zones, a scenario which may allow for the presence of both photospheric and nebular spectral features \citep{Inserra2017, Leloudas2017}.  It is unclear why the appearance of particular nebular features during the photospheric phase seems to occur only in the slowly evolving SLSNe-I.    

We also obtained a spectrum of PS16aqv at $+272$ rest-frame days using GMOS-N with the goal of detecting nebular emission lines.  The spectrum, shown in Figure \ref{gemspec}, is clearly dominated by host galaxy light.  A week after obtaining this spectrum, we obtained deep imaging of PS16aqv in which the SN was not detected to 3$\sigma$ limits of $r>25.6$ and $i>25.3$ mag.  In Figure \ref{gemspec} we also show the nebular spectrum of SN\,2015bn normalized to the $r$-band upper limit.  Assuming the intrinsic spectrum of PS16aqv is well represented by SN\,2015bn, we can clearly see that even the strong nebular emission lines are well below the host galaxy continuum. To test this further, we also plot the spectrum resulting from adding the scaled SN\,2015bn spectrum to the GMOS-N spectrum of PS16aqv.  A few prominent nebular emission lines seen in SN\,2015bn add flux slightly above the level of the noise in the GMOS-N spectrum which may indicate that at $+272$ days PS16aqv has not developed lines as strong as those in SN\,2015bn or that PS16aqv was simply much fainter than the upper limit.  The strongest emission line seen in SN\,2015bn, [\ion{O}{1}]$\lambda6300,6364$, coincides with a strong telluric absorption feature at the redshift of PS16aqv, complicating the identification of a weak emission line.  We consider the GMOS-N spectrum to be representative of the host galaxy spectrum of PS16aqv.

\begin{figure*}
\begin{center}
\includegraphics[scale=0.59]{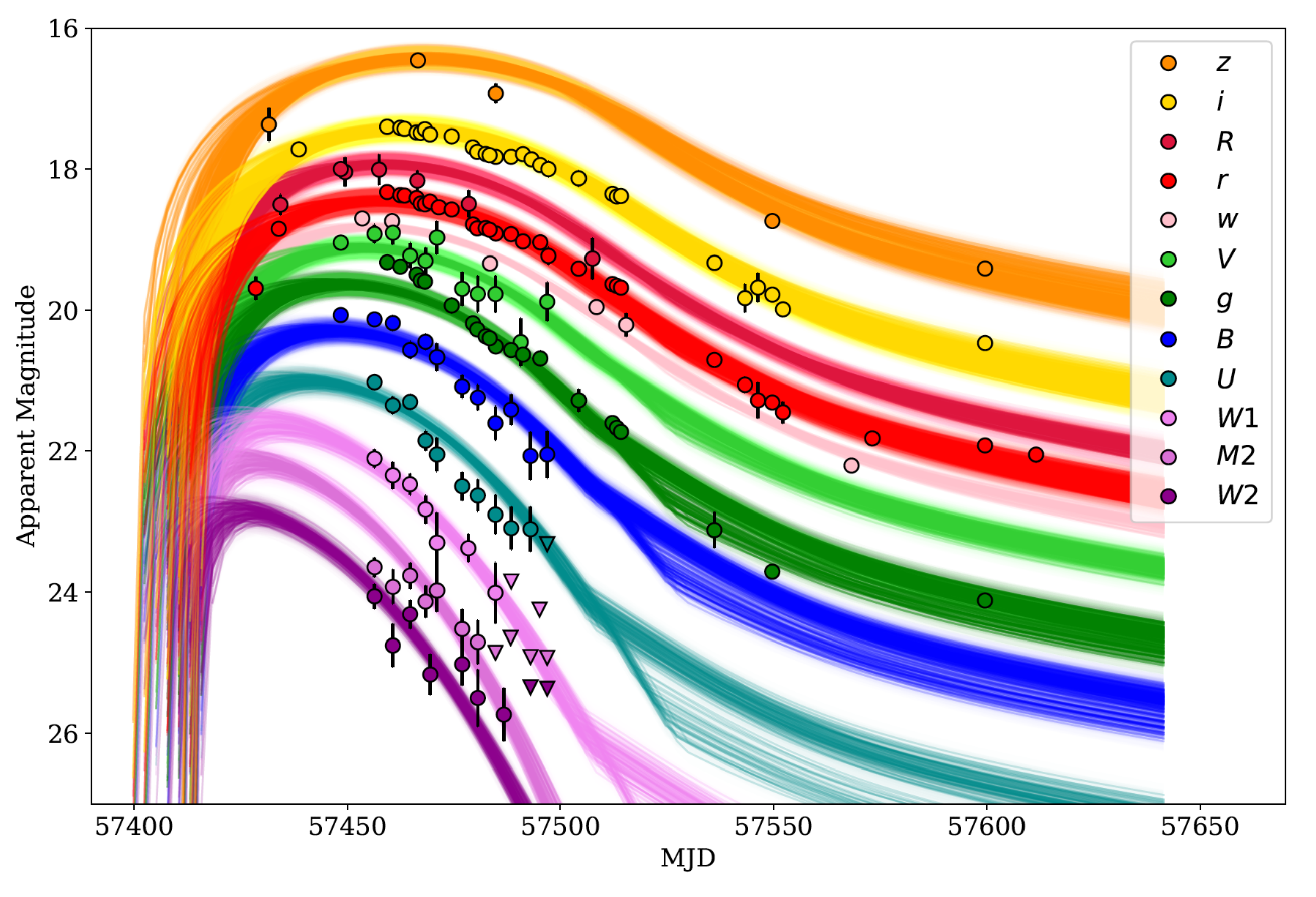}
\end{center}
\caption{Ensemble of magnetar model realizations from our MCMC modeling of PS16aqv with {\tt MOSFiT} compared to the observed data.  The model provides a good overall fit to the trends in the data and the magnetar engine parameters we find (Table \ref{tab:param}) are reasonable compared to the SLSNe-I sample parameters found by \citet{Nicholl2017}, though we find a notably short spin period.  The model favors an internal host extinction value of $A_{V} = 0.55$, relatively high compared to those inferred by \citet{Nicholl2017} and measured from host galaxy observations by \citet{Lunnan2014}. }
\label{magfit}
\end{figure*}

\begin{figure*}
\begin{center}
\includegraphics[scale=0.32]{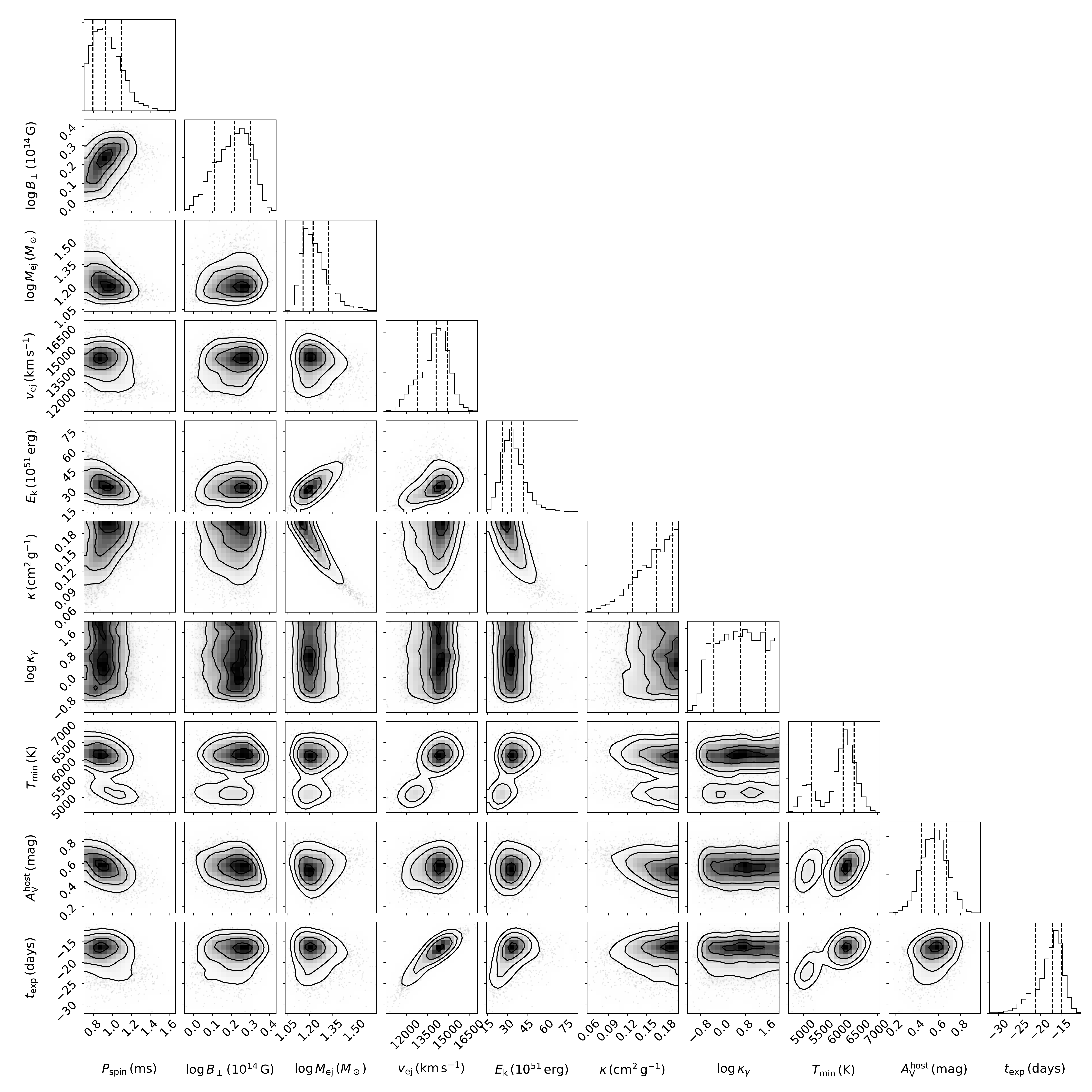}
\end{center}
\caption{Corner plot of the parameter posterior distributions corresponding to the model realizations shown in Figure \ref{magfit}.  The median values and $+/- 1\sigma$ ranges are given in Table \ref{tab:param}.}
\label{magfitcorner}
\end{figure*}

\section{Magnetar Model Fits to PS16aqv}
\label{sec:mag}

Due to its success at explaining the observed lightcurve properties of SLSNe-I \citep{Inserra2013,Nicholl2017}, we use the magnetar central engine model to fit the lightcurves of PS16aqv, and to compare the results with the broad sample.  As with other SLSNe-I, the spectrum of PS16aqv shows no evidence of significant low-velocity CSM.  In addition, the blue spectra and overall lightcurve timescale are inconsistent with $^{56}$Ni decay in a pair-instability SNe \citep{Dessart2012,Mazzali2016}.  Here we use {\tt MOSFiT}, an MCMC code developed specifically for modeling transients \citep{Guillochon2017}.

In {\tt MOSFiT} the model luminosity is calculated using the magnetar engine model as the input power source \citep{KasenBildsten2010}.  The energy input from the spin-down luminosity is fed through an Arnett diffusion model to determine the model bolometric luminosity.  A model for the photosphere is then used to calculate multi-band model magnitudes to be used to fit the observational data.  Following \citet{Nicholl2017} we use a photosphere model that initially expands and cools until reaching a constant temperature, employed to help match the observed temperature evolution of PS16aqv (see Figure \ref{bolLC}).  The SED used to calculate the model magnitudes is a blackbody with a cutoff at 3000 \AA\ used to account for the observed UV absorption in SLSNe-I.  In addition, we constrain the kinetic energy to be less than the total energy available and penalize models which become optically thin in less than 100 days.  We include the same priors as \citet{Nicholl2017} on the resulting 11 free parameters (defined in Table \ref{tab:param}), with the exception of a broader prior on host extinction.

In Figure \ref{magfit} we show an ensemble of multi-band light curve fits to the observations of PS16aqv, and in Figure \ref{magfitcorner} we show the resulting parameter posterior distributions.  The median values of the key engine and ejecta parameters are given in Table \ref{tab:param}.  All of the parameter values fall within the ranges inferred for the sample studied by \citet{Nicholl2017}.  The magnetar model provides a good fit to the overall trend in the data and is able to somewhat reproduce the flattening of the decline rate at about $\sim80$ rest-frame days after peak where the temperature plateaus at around 5000 K and the photosphere begins to recede (see Figure \ref{bolLC}), although the data suggest a more abrupt change in decline rate.  While the bulk lightcurve behavior is well represented by the model, as expected the model cannot account for the undulation (e.g. $g$ and $u$ bands).  Extrapolating the model fits to infer the expected brightness at the epoch of our late limits shown in Figure \ref{obsLC}, we find that the model over-predicts the flux at this time.  To investigate this further we performed another fit including the $r$- and $i$-band upper limits.  The resulting fits near peak are similar but the late-time decline is slightly steeper and so the only differences in the resulting parameters is a slightly weaker magnetic field and lower gamma-ray opacity.  However, the model still over-predicts the flux at the upper limits because this simple model is unable to simultaneously account for the flattening in the lightcurve decline and the late-time limits.            

The inferred $B$-field is moderately strong compared to other SLSNe-I and the spin period is one of the shortest inferred values compared to the full sample distribution \citep{Nicholl2017}, indicating a large reservoir of rotational energy.  In addition, the model prefers a fairly large ejecta mass.  The fast spin and relatively strong $B$-field indicate a fast spin-down time of the magnetar of about 1.7 days.  As \citet{Nicholl2017} point out, the problems associated with powering the observed lightcurves with magnetars that spin down rapidly may be overcome by the fact that the available rotational energy is larger for short spin periods.  The fast spin-down time may explain the overall fast lightcurve decline and temperature evolution despite a relatively high ejecta mass.  Furthermore, the high ejecta mass would delay the onset of the nebular phase which is supported by the slow spectroscopic evolution (see Figure \ref{spec}).  The diverse lightcurve timescale and spectroscopic properties of SLSNe-I may be explained by events with properties located in different regions of ejecta-magnetar parameter space.  
  
Finally, from the model fitting we infer an internal host extinction of $A_{V} = 0.55^{+0.13}_{-0.11}$ mag, a fairly large value compared to measured extinction values from SLSN-I host galaxy studies \citep{Lunnan2014} and the inferred extinction values from the model fitting of the sample in \citet{Nicholl2017}.                 

\capstartfalse
\begin{deluxetable}{cc}[h!]
\tablecolumns{2}
\tabcolsep0.1in\footnotesize
\tablewidth{3in}
\tablecaption{Model parameter medians and 1$\sigma$ ranges corresponding to the posteriors in Figure \ref{magfitcorner} associated with the fits shown in Figure \ref{magfit}  
\label{tab:param}}
\tablehead {
\colhead {Parameter}   &
\colhead {Value}               
}   
\startdata
$P_{\rm spin}$ (ms) & 0.93$^{+0.17}_{-0.18}$  \\[5pt]
log($B$/10$^{14}$ G) & $0.19^{+0.10}_{-0.11}$ \\[5pt]
log($M_{\rm ej}$/M$_{\odot}$) & 1.22$^{+0.09}_{-0.06}$ \\[5pt] 
$v_{\rm ej}$ (km s$^{-1}$) & 14200$^{+700}_{-1400}$  \\[5pt] 
$E_{\rm k}$ (10$^{51}$ erg) & 33.00$^{+10.94}_{-6.18}$ \\[5pt]
$\kappa$ (cm$^{2}$ g$^{-1}$) & 0.16$^{+0.02}_{-0.03}$ \\[5pt]
log $\!\kappa_{\gamma}$ & 0.76$^{+0.80}_{-1.05}$ \\[5pt]
$M_{\rm NS}$ (M$_{\odot}$) & 1.81$^{+0.26}_{-0.31}$ \\[5pt]
$T_{\rm min}$ (K) & 6064$^{+245}_{-948}$ \\[5pt]
$A_{\rm V}^{\rm host}$ & 0.55$^{+0.13}_{-0.11}$ \\[5pt]
$t_{\rm exp}$ (days) & -16.94$^{+2.35}_{-4.71}$ \\[5pt] 
log $\!\sigma$ & -0.83$^{+0.04}_{-0.05}$
\enddata
\tablecomments{$P_{\rm spin}$ is the initial spin period of the magnetar, $B$ is the component of the magnetar magnetic field perpendicular to the spin axis, $M_{\rm ej}$ is the ejecta mass, $v_{\rm ej}$ is the ejecta velocity, $E_{\rm k}$ is the kinetic energy, $\kappa$ is the opacity, $\!\kappa_{\gamma}$ is the gamma-ray opacity, $M_{\rm NS}$ is the neutron star mass, $T_{\rm min}$ is the photosphere temperature floor (described in the text), $A_{\rm V}^{\rm host}$ is the internal host galaxy extinction, $t_{\rm exp}$ is the explosion time relative to the first observation, and $\!\sigma$ is the uncertainty required to yield a reduced chi-squared of 1.  For more details on the model and these parameters see \citet{Nicholl2017}.} 
\end{deluxetable}
\capstarttrue

\begin{figure}
\begin{center}
\includegraphics[scale=0.5]{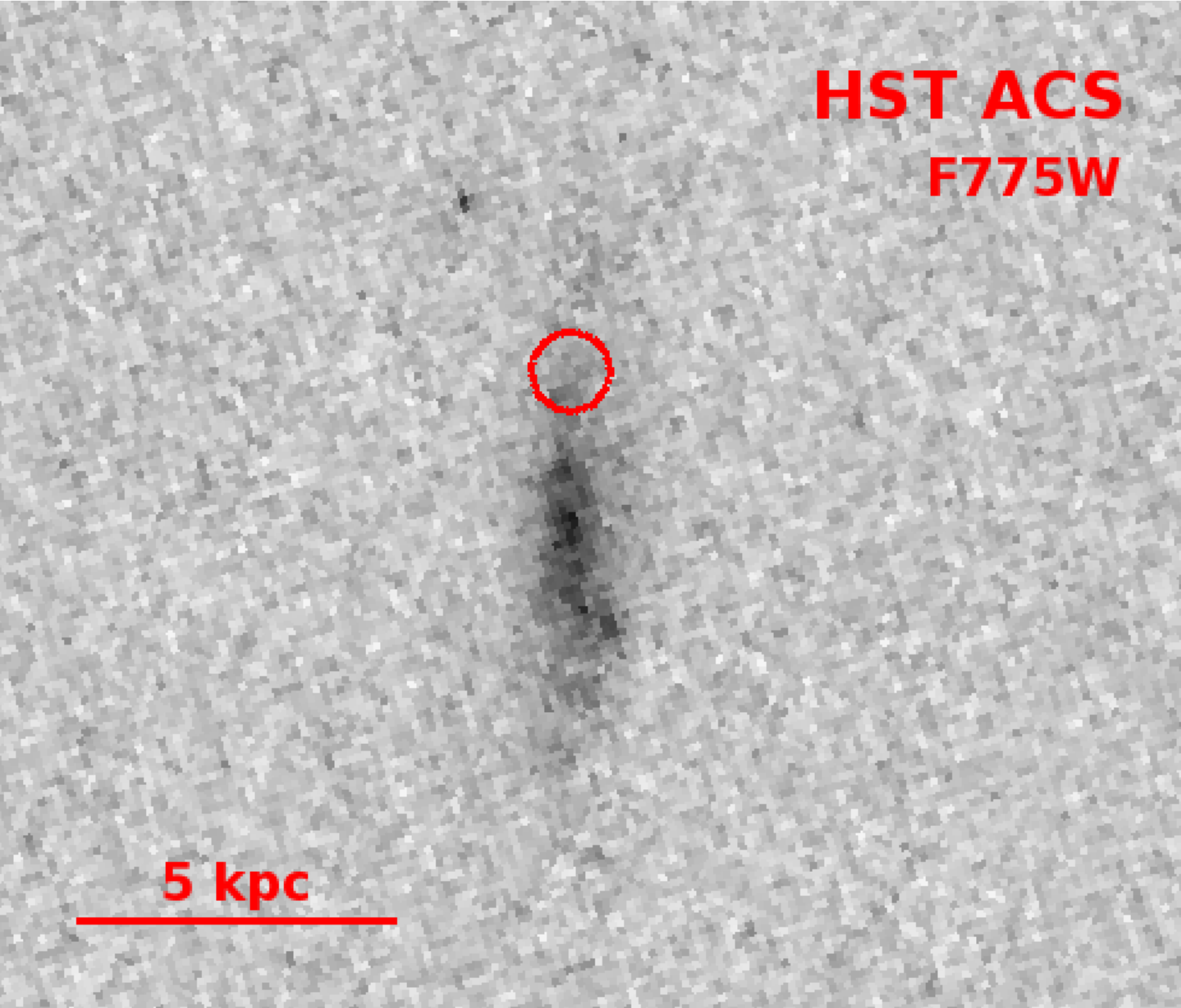}
\end{center}
\caption{\textit{HST} ACS/F775W image of the host galaxy of PS16aqv with the transient location marked (circle; 3$\sigma$).  North is up and East is to the left.  PS16aqv exploded in the outskirts of its host galaxy, offset from the brightest star-forming regions.}
\label{HST}
\end{figure}

\section{Host Galaxy and Environment of PS16aqv}
\label{sec:host}

We measure the $griz$ magnitudes of the host galaxy of PS16aqv from our template images obtained in July 2017 and March 2018.  Using Kron apertures implemented by the {\tt MAG\_AUTO} parameter in {\tt SExtractor} \citep{Bertin1996} we find the following values (corrected for Galactic extinction): $g = 22.70 \pm 0.07$, $r = 22.59 \pm 0.04$, $i = 22.14 \pm 0.05$, and $z = 22.55 \pm 0.15$.  Using the color transformations of \citet{Jordi2006}, we find an absolute $B$-band magnitude of $M_{B} \approx -16.9$, similar to the median value found for the $z \lesssim 0.5$ host sample presented in \citet{Lunnan2014}.  

In Figure \ref{HST} we show our \textit{HST} ACS/F775W image of the host galaxy of PS16aqv showing the location of PS16aqv.  Using the transient and host galaxy centroids, measured with {\tt SExtractor}, we calculate an offset of $R = 0.71 \pm 0.06$ arcseconds, or $2.46 \pm 0.21$  kpc, where the uncertainty is dominated by the astrometric tie uncertainty.  As can be seen in Figure \ref{HST}, PS16aqv occurred in the outskirts of its host galaxy far from the central bright star-forming regions.  Using {\tt SExtractor} we measure the half-light radius of the host galaxy in the \textit{HST} image and find $R_{50} \approx 0.43$ arcseconds, or 1.49 kpc, indicating a compact galaxy similar to other SLSN-I hosts \citep{Lunnan2015}.  This yields a host normalized offset of $R/R_{50} = 1.65$, a larger offset than 87\% of SLSNe-I in the sample studied by \citet{Lunnan2015}.  Following the methodology of \citet{Blanchard2016} we also measure the fractional flux \citep{Fruchter2006}, the fraction of the total galaxy flux coming from pixels fainter than the brightness at the location of PS16aqv, and find a value of $\approx 30\%$, indicating PS16aqv occurred on a relatively faint region of its host galaxy.  This is lower than the values for 75\% of the sample in \citet{Lunnan2015} and is consistent with the large measured offset.  

We use emission lines present in the Gemini spectrum obtained at $+272$ days, which contains negligible SN light, to measure the star formation rate and internal extinction of the host galaxy.  We do not make corrections for underlying stellar absorption.  By measuring the Balmer decrement and assuming an intrinsic value of 2.86 for the ratio of H$\alpha$ to H$\beta$ emission line flux \citep[Case B recombination;][]{Osterbrock1989}, we find a relatively large extinction of $A_{V} = 1.5$ mag.  We note that there is considerable variation in the emission line fluxes along the spatial direction of the 2D spectrum (the slit was aligned along the major-axis of the galaxy; see Figure \ref{gemspec}).  We therefore also measure the Balmer decrement along the precise line of sight to PS16aqv using the $+81$ day LDSS3c spectrum, which contains detections of host lines, and find H$\alpha$/H$\beta$ $= 2.3 \pm 0.7$ which is consistent with no or at most modest extinction.  Several other lines of evidence suggest non-negligible extinction along the line of sight to PS16aqv.  In addition to the inferred extinction of $A_{V} = 0.55$ mag from the model fitting in Section \ref{sec:mag}, a comparison of the spectral shape of PS16aqv with LSQ12dlf at the same phase also indicates extinction.  At $\sim$ 1 month after peak brightness, PS16aqv exhibits a redder spectrum than LSQ12dlf, but applying $A_{V} = 0.5$ mag to LSQ12dlf yields a good match to the spectral shape of PS16aqv.  Given the difference between the global extinction inferred from the Gemini spectrum and the extinction along the line of sight to PS16aqv, we conclude there must be variation in the spatial dust distribution in the galaxy.    

We measure the global star formation rate from the Gemini spectrum using the reddening corrected flux of H$\alpha$ (using $A_{V} = 1.5$ mag) and the SFR calibration of \citet{Kennicutt1998}, yielding SFR = 0.85 M$_{\Sun}$ yr$^{-1}$.  While the global SFR is consistent with that observed for other SLSN-I host galaxies, the variation of the H$\alpha$ flux along the galaxy as apparent in the 2D gemini spectrum indicates a gradient in the SFR.  At the location of PS16aqv we find SFR =  0.16 M$_{\Sun}$ yr$^{-1}$, lower than the central star forming regions.  

Finally, we measure the metallicity using the double-valued $R_{23}$ diagnostic.  We use the $+81$ day LDSS3c spectrum which extends to sufficiently blue wavelengths to measure [\ion{O}{2}] $\lambda3727$, allowing the calculation of $R_{23}$ to determine the metallicity along the line of sight to PS16aqv.  For the lower and upper metallicity branch we find 12 + log(O/H) = 8.1 and 8.5, respectively, using emission line fluxes corrected for a host extinction of $A_{V} = 0.55$ mag.  While we do not detect [\ion{O}{2}] $\lambda4363$ or [\ion{N}{2}] $\lambda6584$, the measured limit of [\ion{N}{2}]/H$\alpha < 0.05$ is sufficiently constraining to rule out the the high metallicity branch.  A metallicity of 12 + log(O/H) = 8.1 is consistent with the range found for the SLSN-I host galaxy population \citep{Lunnan2014}.

\section{Discussion}

\subsection{Lightcurve Undulations}

The physical mechanism responsible for lightcurve undulations remains unknown.  They could be the result of variable engine activity or related to the structure of the ejecta or environment.  Both PS16aqv and SN\,2015bn demonstrate the importance of dense lightcurve time sampling to capture undulations.  Like in SN\,2015bn, the lightcurve undulation in PS16aqv corresponds to when the temperature decline abruptly slows and when the photospheric radius begins to decrease ($\approx$30 rest-frame days after peak), implying the beginning of the recession of the photosphere into the ejecta.  \citet{Nicholl2016} suggested that the temperature change and lightcurve undulation observed in SN\,2015bn may be a signature of the influence of the magnetar wind on the structure of the ejecta.  In particular, \citet{KasenBildsten2010} predicted that the ejecta is swept up into a dense shell with a sharp increase in temperature interior of the shell.  The lightcurve undulation may then be the result of the photosphere reaching the hotter region.

The magnetar may also influence the ejecta by driving ionization fronts \citep{Metzger2014}, which could cause changes in the continuum opacity.  The increased opacity due to the increased ionization would then cause a delay in the escape of radiation, resulting in a change in the lightcurve decline rate.  Finally, the magnetar engine may exhibit flare activity, resulting in intermittent energy injection \citep{YuLi2017}.  

Rather than originate from the power source, the lightcurve undulation may also be the result of interaction with a low-mass CSM ejected by the progenitor star before the explosion.  This could in principle occur even if CSM interaction is not the dominant power source of the lightcurve.  The CSM mass required to power the undulation in PS16aqv is $M_{\rm CSM} \lesssim 0.01$ M$_{\odot}$, similar to the masses inferred for undulations in other events \citep{Nicholl2016,Yan2017,Inserra2017}.  However, the spectrum shows little change during the undulations, making it difficult to disentangle the above scenarios.

In addition to PS16aqv and SN\,2015bn (and possibly LSQ12dlf), there are several other SLSNe-I in the literature which show undulations: SSS120810 \citep{Nicholl2014}, iPTF15esb \citep{Yan2017}, LSQ14an \citep{Inserra2017}, and SN\,2007bi \citep{Gal-Yam2009,Inserra2017}.  While LSQ14an lacks data earlier than $\sim60$ days preventing a comparison with the strongest undulation in SN\,2015bn, \citet{Inserra2017} show that the two events exhibit similar lower amplitude undulations around $+75$ days. The undulations in iPTF15esb show a complex morphology with multiple distinct peaks and the event also shows the emergence of late-time H$\alpha$ emission indicating interaction with neutral H shells \citep{Yan2017}.  The spectroscopic evidence for late-time interaction lends plausibility to the idea that the lightcurve undulations are also caused by interaction.  As in PS16aqv and SN\,2015bn, the undulations in iPTF15esb are stronger in bluer bands.  Though the undulations in iPTF15esb are more significant, SN\,2015bn also shows multiple undulations, in particular a "shoulder" feature before peak and the two "knees" during the decline \citep{Nicholl2016}.  Unfortunately, the sparse time sampling before peak in PS16aqv prevents a comparison.    

In addition to the lightcurve undulation at 30 days post-peak, PS16aqv shows a significant flattening in its decline rate about 80 days after peak. The shallower decline is consistent with the decay of fully trapped $^{56}$Co over the 50 days for which PS16aqv remained observable. However, as shown in Figure \ref{obsLC}, our deep upper limits at $\approx280$ rest-frame days after peak show that this slow decline is clearly not sustained.  At some point during the gap in observations, PS16aqv must have resumed a much faster decline.  This indicates that the flattening at 80 days is more likely related to the ejecta structure or the explosion environment than to $^{56}$Co decay.  Moreover, the flattening corresponds to the time at which the inferred blackbody temperature reaches a plateau. We therefore speculate that it could perhaps be related to abrupt changes in opacity, either due to recombination or the breakout of ionization fronts powered by a magnetar.  Depending on when the lightcurve resumed a faster decline, the dramatic transition at 80 days may be a more pronounced undulation similar to that observed at 30 days or it may be a longer lived ``plateau'' followed by a rapid drop-off.

PS16aqv stands out as a well-observed fast declining SLSN-I with clear evidence for lightcurve undulations similar to those observed in the slow events.  \citet{Inserra2017} investigated three fast declining SLSNe-I and found no clear evidence for undulations (though pointed out a possible undulation in LSQ12dlf), suggesting that lightcurve undulations only occur in slowly evolving SLSNe-I such as SN\,2015bn.  PS16aqv is a clear counterexample and lends additional support to the idea that there is a single class of SLSNe-I with a consistent explosion mechanism but with varying ejecta/engine properties.  Early samples indicated a possible bimodality in timescales \citep{Nicholl2015} but recent larger sample studies suggest that SLSNe-I form a continuum of timescales rather than two distinct fast and slow groups \citep{Nicholl2017, DeCia2017, Lunnan2018}.  In addition, \citet{Nicholl2017} show that the engine parameter distributions of fast and slow SLSNe-I overlap with no clear offset; the slow events simply prefer somewhat lower magnetic fields and higher ejecta masses.  Observing undulations across the range of lightcurve timescales supports a uniform origin.

\subsection{Limits on Radioactive Ejecta}
  
Finally, we use the late-time observations to place a limit on the cobalt mass, $M_{\rm Co}$, since any luminosity from $^{56}$Co decay must be lower than the measured upper limits.  Using the standard equation for energy injection by radioactive decay of $^{56}$Co assuming full gamma-ray trapping, we find a limit of $M_{\rm Co} \lesssim 0.35$ M$_{\odot}$.  As inferred for previous SLSNe \citep{Pastorello2010,Inserra2013}, this implies a $^{56}$Ni mass far below that required to explain the peak lightcurve luminosity with radioactive decay alone ($\approx\!$ 28 M$_{\odot}$).  Our limit on the $^{56}$Ni mass is lower than masses inferred from the late-time decline phase of other SLSNe-I \citep[$\approx1-4$ M$_{\odot}$;][]{Inserra2013} under the same assumption of full gamma-ray trapping, making it the most stringent constraint on radioactive decay in SLSNe-I.  In fact, our deep limit indicates a synthesized $^{56}$Ni mass lower than that inferred for some energetic Type Ic SNe \citep[e.g.~SN\,1998bw;][]{Sollerman2002}, and therefore suggests that SLSNe-I do not produce larger $^{56}$Ni masses than energetic Type Ic SNe.  

An important caveat is that gamma-rays are expected to leak out of the ejecta as the optical depth decreases \citep{Sollerman2004}.  Over time the energy deposition provided by the kinetic energy of positrons, which is about 3.4\% of the total released energy, becomes the dominant source of energy from radioactive decay.  Under a somewhat pessimistic assumption that the optical depth to gamma-rays reaches unity by about 50 days after peak, the limit on the cobalt mass implied by the upper limit becomes $M_{\rm Co} \lesssim 5$ M$_{\odot}$, still much lower than the $^{56}$Ni mass required to power the peak luminosity.  As this would roughly affect all SLSNe-I equally, this does not change the observation that \textit{compared to other SLSNe-I}, the deep limit on PS16aqv's late-time luminosity implies a low $^{56}$Ni mass.  This observation supports a picture in which at least some SLSNe-I do not produce significantly more $^{56}$Ni than typical core-collapse SNe, the primary difference being the presence of a central engine in SLSNe-I.                     
      
\section{Conclusions}
We present an extensive photometric and spectroscopic dataset from ground- and space-based telescopes for the SLSN-I PS16aqv.  While the photospheric spectra and overall lightcurve evolution timescale are most similar to fast-declining SLSNe-I, PS16aqv shows a remarkably similar lightcurve undulation at $30$ days after peak as the well-studied slowly evolving SLSN-I SN\,2015bn.  Well-observed undulations have previously only been seen in the slower evolving SLSNe-I.  While the physical mechanism of lightcurve undulations in SLSNe-I remains unknown, it is likely related to either engine activity or the structure of the ejecta or environment.  The presence of undulations in SLSNe-I with a range of decline rates lends support to the notion that fast and slow SLSNe-I share the same explosion mechanism and that they are linked by a continuum of engine/ejecta properties.  The distributions of these properties may naturally explain other unusual SLSN-I properties, such as fast lightcurve evolution coupled with slow spectroscopic evolution (as observed in PS16aqv) which may be due to a fast magnetar spin-down time coupled with high ejecta mass.       

In addition, deep late-time limits after PS16aqv settled on to a very slow decline phase suggest that it may have exhibited another more pronounced undulation starting at $+80$ days or it may have exhibited a long-lived plateau before rapidly fading.  The growing number of SLSNe-I like PS16aqv with lightcurve complexity highlights the importance of obtaining well-sampled lightcurves of future events.  Identifying the origin of lightcurve undulations requires large samples of well-observed events in order to search for potential correlations between undulation characteristics and ejecta/engine properties.  The late-time limits also yielded a tight constraint on the synthesized nickel mass ($M_{\rm Ni} \lesssim 0.35$ M$_{\odot}$), lower than estimates from other SLSNe-I. 

Using deep \textit{HST} imaging and late-time Gemini spectroscopy we also studied the host galaxy of PS16aqv.  A spatially resolved host spectrum indicates a spatially varying extinction and star formation rate, with the explosion site located in a faint region 2.46 kpc from the central bright region which corresponds to a large host-normalized offset.  While the global host extinction is large ($A_{V}  \approx 1.5$ mag), the value inferred along the line of site to PS16aqv from our lightcurve modeling is more modest ($A_{V}  \approx 0.55$ mag), though both results suggest the host galaxy of PS16aqv has high extinction compared to other SLSN-I hosts.  The rather unremarkable host location of PS16aqv motivates further study into the question of whether the sub-galactic locations of SLSNe-I show a strong preference for bright regions of their hosts, like long GRBs.  Increasing the sample size of SLSNe-I with high-resolution host galaxy observations is key to making progress in our understanding of their environments.

\acknowledgments
The Berger Time-Domain Group at Harvard is supported in part by the NSF under grant AST-1714498 and by NASA under grant NNX15AE50G.  This paper is based upon work supported by the National Science Foundation Graduate Research Fellowship Program under Grant No. DGE1144152.  We thank Pete Challis and Allyson Bieryla for assistance with some of the FLWO 48-inch observations.  We thank Stephen Smartt and Ken Smith for providing access to the early PSST images.  This work is based in part on observations obtained at the MDM Observatory, operated by Dartmouth College, Columbia University, Ohio State University, Ohio University, and the University of Michigan.  This work is partially based on data acquired with the Swift GO program 1114109 (PI Margutti).  R. M. acknowledges partial support from programs No. NNX16AT51G provided by NASA through Swift Guest Investigator Programs.  Based on observations (Proposal ID GN-2016B-FT-28) obtained at the Gemini Observatory acquired through the Gemini Observatory Archive and processed using the Gemini IRAF package, which is operated by the Association of Universities for Research in Astronomy, Inc., under a cooperative agreement with the NSF on behalf of the Gemini partnership: the National Science Foundation (United States), the National Research Council (Canada), CONICYT (Chile), Ministerio de Ciencia, Tecnolog\'{i}a e Innovaci\'{o}n Productiva (Argentina), and Minist\'{e}rio da Ci\^{e}ncia, Tecnologia e Inova\c{c}\~{a}o (Brazil).  This paper uses data products produced by the OIR Telescope Data 
Center, supported by the Smithsonian Astrophysical Observatory.  Some observations reported here were obtained at the MMT Observatory, a joint facility of the Smithsonian Institution and the University of Arizona.  This paper includes data gathered with the 6.5 meter Magellan Telescopes located at Las Campanas Observatory, Chile.  Based on observations made with the NASA/ESA Hubble Space Telescope, obtained from the data archive at the Space Telescope Science Institute.  Support for program GO-15162 was provided by NASA through a grant from the Space Telescope Science Institute, which is operated by the Association of Universities for Research in Astronomy, Inc., under NASA contract NAS 5-26555.

\bibliographystyle{apj}

\appendix
\section{Photometry}

\capstartfalse
\begin{deluxetable*}{cccccccccc}[!b]
\tablecolumns{10}
\tabcolsep0.1in\footnotesize
\tablewidth{7in}
\tablecaption{Ground-Based Observations of PS16aqv ($wgriz$ are in AB magnitudes and $BVR$ and CSS are in Vega magnitudes) 
\label{tab:ground}}
\tablehead {
\colhead {MJD}   &
\colhead {$B$}     &
\colhead {$V$} &
\colhead {$g$} &
\colhead{$r$}  &
\colhead {$R$}   &
\colhead {$CSS$} &
\colhead{$w$} &
\colhead{$i$} &
\colhead{$z$}           
}   
\startdata
57428.59 & \nodata & \nodata & \nodata & 19.69 (0.16) & \nodata & \nodata & \nodata & \nodata & \nodata \\
57431.67 & \nodata & \nodata & \nodata & \nodata & \nodata & \nodata & \nodata & \nodata & 19.37 (0.23) \\
57434.00 & \nodata & \nodata & \nodata & 18.85 (0.02) & \nodata & \nodata & \nodata & \nodata & \nodata \\
57434.40 & \nodata & \nodata & \nodata & \nodata & \nodata & 18.81 (0.14) & \nodata & \nodata & \nodata \\
57438.62 & \nodata & \nodata & \nodata & \nodata & \nodata & \nodata & \nodata & 18.72 (0.02) & \nodata \\
57448.48 & 18.57 (0.08) & 18.45 (0.04) & \nodata & \nodata & 18.30 (0.06) & \nodata & \nodata & \nodata & \nodata \\
57449.50 & \nodata & \nodata & \nodata & \nodata & \nodata & 18.34 (0.20) & \nodata & \nodata & \nodata \\
57453.54 & \nodata & \nodata & \nodata & \nodata & \nodata & \nodata & 18.40 (0.01) & \nodata & \nodata \\
57457.50 & \nodata & \nodata & \nodata & \nodata & \nodata & 18.31 (0.21) & \nodata & \nodata & \nodata \\
57459.39 & \nodata & \nodata & \nodata & \nodata & \nodata & \nodata & \nodata & 18.40 (0.07) & \nodata \\
57459.40 & \nodata & \nodata & 18.32 (0.06) & 18.33 (0.06) & \nodata & \nodata & \nodata & \nodata & \nodata \\
57460.55 & \nodata & \nodata & \nodata & \nodata & \nodata & \nodata & 18.44 (0.01) & \nodata & \nodata \\
57462.48 & \nodata & \nodata & \nodata & 18.37 (0.06) & \nodata & \nodata & \nodata & 18.42 (0.07) & \nodata \\
57462.49 & \nodata & \nodata & 18.38 (0.06) & \nodata & \nodata & \nodata & \nodata & \nodata & \nodata \\
57463.47 & \nodata & \nodata & \nodata & \nodata & \nodata & \nodata & \nodata & 18.43 (0.08) & \nodata \\
57463.48 & \nodata & \nodata & \nodata & 18.38 (0.06) & \nodata & \nodata & \nodata & \nodata & \nodata \\
57466.32 & \nodata & \nodata & 18.50 (0.08) & 18.41 (0.07) & \nodata & \nodata & \nodata & 18.48 (0.06) & \nodata \\
57466.50 & \nodata & \nodata & \nodata & \nodata & \nodata & 18.46 (0.14) & \nodata & \nodata & \nodata \\
57466.64 & \nodata & \nodata & \nodata & \nodata & \nodata & \nodata & \nodata & \nodata & 18.46 (0.02) \\
57467.28 & \nodata & \nodata & \nodata & 18.49 (0.07) & \nodata & \nodata & \nodata & 18.49 (0.07) & \nodata \\
57467.29 & \nodata & \nodata & 18.58 (0.08) & \nodata & \nodata & \nodata & \nodata & \nodata & \nodata \\
57468.28 & \nodata & \nodata & \nodata & 18.50 (0.06) & \nodata & \nodata & \nodata & 18.44 (0.07) & \nodata \\
57468.29 & \nodata & \nodata & 18.59 (0.08) & \nodata & \nodata & \nodata & \nodata & \nodata & \nodata \\
57469.48 & \nodata & \nodata & \nodata & 18.46 (0.06) & \nodata & \nodata & \nodata & 18.51 (0.07) & \nodata \\
57471.51 & \nodata & \nodata & \nodata & 18.54 (0.09) & \nodata & \nodata & \nodata & \nodata & \nodata \\
57474.46 & \nodata & \nodata & \nodata & \nodata & \nodata & \nodata & \nodata & 18.54 (0.08) & \nodata \\
57474.47 & \nodata & \nodata & 18.93 (0.10) & 18.57 (0.07) & \nodata & \nodata & \nodata & \nodata & \nodata \\
57478.50 & \nodata & \nodata & \nodata & \nodata & \nodata & 18.80 (0.18) & \nodata & \nodata & \nodata \\
57479.41 & \nodata & \nodata & \nodata & \nodata & \nodata & \nodata & \nodata & 18.69 (0.08) & \nodata \\
57479.46 & \nodata & \nodata & 19.19 (0.08) & 18.79 (0.06) & \nodata & \nodata & \nodata & \nodata & \nodata \\
57480.46 & \nodata & \nodata & \nodata & 18.84 (0.07) & \nodata & \nodata & \nodata & 18.76 (0.09) & \nodata \\
57480.47 & \nodata & \nodata & 19.28 (0.08) & \nodata & \nodata & \nodata & \nodata & \nodata & \nodata \\
57482.45 & \nodata & \nodata & 19.37 (0.08) & 18.84 (0.07) & \nodata & \nodata & \nodata & 18.78 (0.09) & \nodata \\
57483.38 & \nodata & \nodata & 19.40 (0.08) & 18.86 (0.06) & \nodata & \nodata & \nodata & 18.80 (0.08) & \nodata \\
57483.48 & \nodata & \nodata & \nodata & \nodata & \nodata & \nodata & 19.04 (0.01) & \nodata & \nodata \\
57484.84 & \nodata & \nodata & 19.52 (0.07) & 18.92 (0.07) & \nodata & \nodata & \nodata & 18.83 (0.08) & 18.93 (0.14) \\
57488.43 & \nodata & \nodata & \nodata & 18.93 (0.07) & \nodata & \nodata & \nodata & 18.83 (0.08) & \nodata \\
57488.44 & \nodata & \nodata & 19.57 (0.08) & \nodata & \nodata & \nodata & \nodata & \nodata & \nodata \\
57491.23 & \nodata & \nodata & \nodata & \nodata & \nodata & \nodata & \nodata & 18.79 (0.08) & \nodata \\
57491.24 & \nodata & \nodata & 19.63 (0.09) & 19.03 (0.07) & \nodata & \nodata & \nodata & \nodata & \nodata \\
57493.25 & \nodata & \nodata & \nodata & \nodata & \nodata & \nodata & \nodata & 18.86 (0.08) & \nodata \\
57495.30 & \nodata & \nodata & \nodata & 19.04 (0.08) & \nodata & \nodata & \nodata & 18.94 (0.09) & \nodata \\
57495.31 & \nodata & \nodata & 19.69 (0.10) & \nodata & \nodata & \nodata & \nodata & \nodata & \nodata \\
57497.22 & \nodata & \nodata & \nodata & 19.23 (0.11) & \nodata & \nodata & \nodata & 19.00 (0.11) & \nodata \\
57504.32 & \nodata & \nodata & \nodata & 19.41 (0.08) & \nodata & \nodata & \nodata & 19.13 (0.11) & \nodata \\
57504.33 & \nodata & \nodata & 20.28 (0.15) & \nodata & \nodata & \nodata & \nodata & \nodata & \nodata \\
57507.50 & \nodata & \nodata & \nodata & \nodata & \nodata & 19.57 (0.29) & \nodata & \nodata & \nodata \\
57508.43 & \nodata & \nodata & \nodata & \nodata & \nodata & \nodata & 19.66 (0.01) & \nodata & \nodata \\
57512.18 & \nodata & \nodata & 20.60 (0.07) & 19.63 (0.04) & \nodata & \nodata & \nodata & 19.35 (0.05) & \nodata \\
57513.20 & \nodata & \nodata & \nodata & 19.66 (0.03) & \nodata & \nodata & \nodata & 19.39 (0.05) & \nodata \\
57513.21 & \nodata & \nodata & 20.67 (0.07) & \nodata & \nodata & \nodata & \nodata & \nodata & \nodata \\
57514.17 & \nodata & \nodata & \nodata & \nodata & \nodata & \nodata & \nodata & 19.38 (0.05) & \nodata \\
57514.18 & \nodata & \nodata & 20.73 (0.07) & 19.68 (0.04) & \nodata & \nodata & \nodata & \nodata & \nodata \\
57515.40 & \nodata & \nodata & \nodata & \nodata & \nodata & \nodata & 19.91 (0.16) & \nodata & \nodata \\
57536.16 & \nodata & \nodata & 22.12 (0.25) & 20.71 (0.03) & \nodata & \nodata & \nodata & 20.33 (0.06) & \nodata \\
57543.25 & \nodata & \nodata & \nodata & 21.06 (0.10) & \nodata & \nodata & \nodata & 20.83 (0.20) & \nodata \\
57546.31 & \nodata & \nodata & \nodata & 21.28 (0.25) & \nodata & \nodata & \nodata & 20.68 (0.20) & \nodata \\
57549.68 & \nodata & \nodata & 22.71 (0.06) & 21.31 (0.02) & \nodata & \nodata & \nodata & 20.78 (0.06) & 20.74 (0.07) \\
57552.18 & \nodata & \nodata & \nodata & 21.45 (0.15) & \nodata & \nodata & \nodata & 20.99 (0.10) & \nodata \\
57568.30 & \nodata & \nodata & \nodata & \nodata & \nodata & \nodata & 21.91 (0.07) & \nodata & \nodata \\
57573.18 & \nodata & \nodata & \nodata & 21.82 (0.07) & \nodata & \nodata & \nodata & \nodata & \nodata \\
57599.61 & \nodata & \nodata & 23.12 (0.08) & 21.92 (0.02) & \nodata & \nodata & \nodata & 21.47 (0.07) & 21.41 (0.08) \\
57611.50 & \nodata & \nodata & \nodata & 22.05 (0.10) & \nodata & \nodata & \nodata & \nodata & \nodata \\
57785.20 & \nodata & \nodata & \nodata & \nodata & \nodata & \nodata & \nodata & >25.3 & \nodata \\
57789.50 & \nodata & \nodata & \nodata & >25.6 & \nodata & \nodata & \nodata & \nodata & \nodata 
\enddata
\tablecomments{These magnitudes are not corrected for Galactic extinction.}
\end{deluxetable*}   
\capstarttrue

\capstartfalse
\begin{deluxetable*}{ccccccc}[!htb]
\tablecolumns{7}
\tabcolsep0.1in\footnotesize
\tablewidth{7in}
\tablecaption{\textit{Swft} Observations of PS16aqv (Vega magnitudes) 
\label{tab:swift}}
\tablehead {
\colhead {MJD}   &
\colhead {$UVW2$}     &
\colhead {$UVM2$} &
\colhead {$UVW1$} &
\colhead{$U$}  &
\colhead {$B$}   &
\colhead {$V$}           
}   
\startdata
57456.41 & \nodata & \nodata & 19.11 (0.13) & \nodata & \nodata & \nodata \\
57456.42 & 20.06 (0.17) & \nodata & \nodata & 18.02 (0.08) & 18.63 (0.08) & 18.32 (0.13) \\
57456.43 & \nodata & 20.15 (0.13) & \nodata & \nodata & \nodata & \nodata \\
57460.74 & \nodata & \nodata & 19.34 (0.19) & 18.35 (0.12) & 18.68 (0.11) & \nodata \\
57460.75 & 20.76 (0.30) & 20.43 (0.24) & \nodata & \nodata & \nodata & 18.30 (0.16) \\
57464.76 & \nodata & \nodata & 19.47 (0.15) & \nodata & \nodata & \nodata \\
57464.79 & 20.32 (0.20) & 20.27 (0.18) & \nodata & 18.30 (0.10) & 19.07 (0.12) & 18.63 (0.18) \\
57468.42 & \nodata & \nodata & 19.83 (0.19) & 18.85 (0.14) & 18.95 (0.11) & \nodata \\
57468.43 & \nodata & 20.64 (0.23) & \nodata & \nodata & \nodata & 18.70 (0.18) \\
57469.55 & 21.17 (0.29) & \nodata & \nodata & \nodata & \nodata & \nodata \\
57471.07 & \nodata & \nodata & 20.30 (0.42) & \nodata & \nodata & \nodata \\
57471.08 & \nodata & 20.48 (0.29) & \nodata & 19.05 (0.24) & 19.17 (0.19) & 18.37 (0.23) \\
57476.93 & 21.02 (0.30) & 21.02 (0.28) & \nodata & 19.50 (0.20) & 19.59 (0.16) & 19.10 (0.23) \\
57478.38 & \nodata & \nodata & 20.38 (0.19) & \nodata & \nodata & \nodata \\
57480.60 & \nodata & \nodata & \nodata & 19.63 (0.22) & 19.74 (0.18) & \nodata \\
57480.61 & 21.50 (0.40) & 21.21 (0.31) & \nodata & \nodata & \nodata & 19.17 (0.25) \\
57484.76 & \nodata & \nodata & 21.01 (0.43) & 19.90 (0.27) & 20.10 (0.23) & 19.17 (0.25) \\
57484.77 & \nodata & >21.36 & \nodata & \nodata & \nodata & \nodata \\
57486.73 & 21.74 (0.38) & \nodata & \nodata & \nodata & \nodata & \nodata \\
57488.33 & \nodata & >21.15 & \nodata & \nodata & \nodata & \nodata \\
57488.43 & \nodata & \nodata & \nodata & \nodata & 19.91 (0.21) & \nodata \\
57488.46 & \nodata & \nodata & >20.85 & 20.09 (0.30) & \nodata & \nodata \\
57490.72 & \nodata & \nodata & \nodata & \nodata & \nodata & 19.86 (0.34) \\
57493.00 & \nodata & \nodata & \nodata & 20.11 (0.31) & 20.57 (0.34) & \nodata \\
57493.01 & >21.35 & >21.42 & \nodata & \nodata & \nodata & \nodata \\
57495.14 & \nodata & \nodata & >21.25 & \nodata & \nodata & \nodata \\
57496.93 & >21.37 & >21.43 & \nodata & >20.32 & 20.55 (0.33) & 19.28 (0.27)
\enddata
\tablecomments{These magnitudes are not corrected for Galactic extinction.}
\end{deluxetable*}   
\capstarttrue

\end{document}